\documentclass[journal]{IEEEtran}

\usepackage[utf8]{inputenc}
\usepackage[cmex10]{amsmath}%
\usepackage{amsfonts}%
\usepackage{amssymb}%
\usepackage{graphicx}
\usepackage{balance}
\usepackage{adjustbox}
\usepackage{acronym}
\usepackage{nomencl}
\usepackage{array}
\RequirePackage{ifthen}
\renewcommand{\nomgroup}[1]{%
\ifthenelse{\equal{#1}{S}}{\item[\textbf{Symbols}]}{
\ifthenelse{\equal{#1}{A}}{\item[\textbf{Acronyms}]}{}}}

\usepackage{listings}
\usepackage[left=2.5cm,top=2.5cm,right=2cm,bottom=2cm]{geometry}
\usepackage{array}
\usepackage{amsthm}
\usepackage[english]{babel}
\usepackage{url}
\usepackage{lastpage}
\usepackage{fancyhdr}
\usepackage[dvipsnames,usenames]{color}
\usepackage{ae}
\usepackage{paralist}
\usepackage[T1]{fontenc}
\usepackage{blindtext}
\usepackage{mhchem}
\usepackage{multirow,multicol}
\usepackage{todonotes}
\usepackage{arydshln}
\usepackage{soul}
\usepackage{multirow,multicol}
\usepackage{todonotes}
\usepackage{arydshln}
\usepackage{soul}
\usepackage[normalem]{ulem}
\usepackage{subcaption}
\usepackage{amsmath}
\usepackage{tikz}
\usetikzlibrary{automata,positioning}
\usepackage{colortbl}
\usepackage{forest}
\usepackage{smartdiagram}
\usepackage{listings}
\usepackage[ruled]{algorithm2e}
\usesmartdiagramlibrary{additions} 
\usetikzlibrary{shapes.geometric}

\newcolumntype{P}[1]{>{\centering\arraybackslash}p{#1}}

\makenomenclature

\begin{document}


\title{Model-Based Risk Assessment for Cyber Physical Systems Security}

\author{Ashraf~Tantawy,	Abdelkarim~Erradi, Sherif~Abdelwahed, and~Khaled~Shaban
	\thanks{A. Tantawy and S. Abdelwahed are with the Department
		of Electrical and Computer Engineering, Virginia Commonwealth University, VA, 23220 USA e-mail: (amatantawy, sabdelwahed@vcu.edu)}
\thanks{A. Erradi and K. Shaban are with the Computer Science and Engineering Department, Qatar University, Doha, Qatar e-mail: (erradi, khaled.shaban)@qu.edu.qa}
}

\markboth{}%
{Tantawy \MakeLowercase{\textit{et al.}}}

\maketitle

\begin{abstract}
Traditional techniques for Cyber-Physical Systems (CPS) security design either treat the cyber and physical systems independently, or do not address the specific vulnerabilities of real time embedded controllers and networks used to monitor and control physical processes. In this work, we develop and test an integrated model-based approach for CPS security risk assessment utilizing a CPS testbed with real-world industrial controllers and communication protocols. The testbed monitors and controls an exothermic Continuous Stirred Tank Reactor (CSTR) simulated in real-time. CSTR is a fundamental process unit in many industries, including Oil \& Gas, Petrochemicals, Water treatment, and nuclear industry. In addition, the process is rich in terms of hazardous scenarios that could be triggered by cyber attacks due to the lack of possible mechanical protection. The paper presents an integrated approach to analyze and design the cyber security system for a given CPS where the physical threats are identified first to guide the risk assessment process. A mathematical model is derived for the physical system using a hybrid automaton to enumerate potential hazardous states of the system. The cyber system is then analyzed using network and data flow models to develop the attack scenarios that may lead to the identified hazards. Finally, the attack scenarios are performed on the testbed and observations are obtained on the possible ways to prevent and mitigate the attacks. The insights gained from the experiments result in several key findings, including the expressive power of hybrid automaton in security risk assessment, the hazard development time and its impact on cyber security design, and the tight coupling between the physical and the cyber systems for CPS that requires an integrated design approach to achieve cost-effective and secure designs.
\end{abstract}


\IEEEpeerreviewmaketitle

\section{Introduction} 
Cyber-Physical Systems (CPS) integrate physical elements, for sensing and actuation, with cyber elements, for computation and communication, to automate and control industrial processes. CPS is pervasively used in critical application domains such as health care, traffic management, manufacturing, and energy infrastructures. These systems are increasingly adopting commercial and open source software components and standard communication protocols in order to reduce infrastructure costs and ease integration and connection with corporate networks. However, this has exposed such systems to new security threats and made them a prime target for cyber-attacks to disrupt their normal operation. This may result in profound and catastrophic impacts such as endangering public safety and economic stability. Despite ongoing efforts to secure and protect CPS, these critical infrastructure components remain vulnerable to cyber attacks. Recent intensified sophisticated attacks on these systems have stressed the importance of methodologies and tools to assess and manage cyber security risks \cite{Alcaraz2015}. Additionally, it is necessary to identify and address safety and security requirements earlier as part of the system design process \cite{Friedberg2017}.

Traditional IT security risk assessment is a well-established domain, guided by several international standards, e.g., \cite{Stoneburner2002}. More recently, standards emerged to address the specific needs of CPS domains, such as IEC 62443 for securing industrial automation and control systems. However, these standards provide best practices for the security system independent of the monitored physical system. It is the responsibility of the CPS designer to integrate the safety and security aspects of the CPS, often in an ad-hoc manner. Realizing this gap, IEC/TC65 plenary board proposed a new group to consider how to bridge functional safety and cyber security for industrial automation systems \cite{Kanamaru2017}.

In this work, we propose a model-based design approach where physical system modeling, data flow modeling, and attack trees are integrated to deliver a unified design framework of safe and secure CPS. First, the physical system modeling documents all the system components and their interaction interfaces including  system sensors, controllers, and the supporting networks and protocols. Then we identify the data flow and information exchange between the system components to enable monitoring and controlling the physical process. The resulting system model and data flow model are used as inputs for threat modeling using attack trees. The latter are conceptual diagrams describing the system threats and possible attacks to realize those threats \cite{mauw2005foundations}.

We summarize the key contributions of the paper as follows: \begin{inparaenum}[(1)] \item the introduction of hybrid system automaton as a powerful tool for cyber security risk assessment, and countermeasure \& mitigation design. \item Development of an integrated safety-security model-based co-design approach for CPS. The approach integrates physical and cyber models for attack scenario generation, risk assessment, and countermeasures design. \item Providing insight on key research directions as revealed throughout the design and implementation of the case study CPS.\end{inparaenum}

We applied and empirically validated our proposed model-based approach for CPS security risk assessment on a CPS testbed that monitors and controls an exothermic Continuous Stirred Tank Reactor (CSTR) simulated in real-time. CSTR is a widely used model for chemical reactor engineering. This process is selected because of its practical importance and its associated hazards that can be triggered by cyber attacks with no possible mechanical protection, including reactor overflow and thermal runaway with the risk of fire, explosion or environmental hazards \cite{Friedberg2017}.
A technical report detailing the design of the CPS testbed along with the implementation of the CSTR simulator, the Basic Process Control System (BPCS) and the Safety Instrumented System (SIS) are available online \footnote{\url{https://github.com/qucse/CpsSecurity}}.

The rest of the paper is organized as follows: Section \ref{sec:Related Work} discusses important related work. Section \ref{sec:CPS Description} presents the overall architecture for the CPS used as a case study throughout the paper. Section \ref{sec:hazard-id} introduces the hybrid automaton approach for hazard identification. The design of cyber attacks is presented in Section \ref{sec:cyber-attacks}. Section \ref{sec:pentest} summarizes the penetration testing results and the formal risk assessment. The key insights gained from the case study are summarized in Section \ref{sec:discussion}, and the paper is concluded with future research directions in Section \ref{sec:Conclusion}.

\section{Related Work} \label{sec:Related Work}
In addition to international standards, the research community recently started to address the CPS security problem in non-traditional ways. A survey on CPS security, challenges, and solutions could be found in \cite{Ashibani2017}. A review of risk assessment methods for SCADA systems is conducted in \cite{Cherdantseva2016}, including attack trees, countermeasure trees, petri nets, as well as several quantitative risk measures. CPS security for industrial processes has been studied in \cite{Huang2015}, where a multi-layer cyber-security protection architecture is proposed. A system for runtime attack detection and prevention for industrial control systems is proposed in \cite{Ylmaz2018}. The application of attack-defense trees to analyze cyber security for CPS is reported in \cite{Ji2016}. The authors in \cite{Sabaliauskaite2015} proposed an integrated CPS safety and security lifecycle process, merging ISA84/IEC 61511 and ISA99/IEC 62443 lifecycle processes, where a combined failure and attack graph is proposed for risk assessment. A safety/security risk analysis approach that combines bowtie analysis for safety systems with attack tree analysis for security systems is considered in \cite{Abdo2018a}. CPS security for the electric power grid is discussed in \cite{Sridhar2012}, where the authors proposed a risk assessment methodology and identified the potential threats for each component of the grid.

As a research aid, researchers build testbeds that represent scaled down versions of actual physical industrial control systems in order to provide a research environment that allows extensive attack-defense experimentation with realistic attack scenarios. The testbed facilitates the vulnerability and risk assessment, impact analysis of cyber-attacks on the controlled process, and risk mitigation, in order to enable the design and evaluation of effective detection and defense mechanisms. Several testbeds developed at various universities and national research labs have been reported in the literature for different CPS application domains, predominantly focusing on critical infrastructures such as the smart power grid testbeds presented in \cite{Hahn2013}, \cite{Poudel2017}, \cite{Jarmakiewicz2017}. One of the early power grid testbeds are the National SCADA TestBed (NSTB) \cite{Craig2008} and Idaho National Labs (INL) SCADA Testbed which deploys real physical grid components including generators, transmitters, substations and controllers \cite{Inl2009}. In addition to the energy domain, a smart transportation system testbed has been proposed in \cite{koutsoukos2017sure}. Industrial Control Systems SCADA testbed has been discussed in \cite{Morris2011}, and a water treatment testbed is reported in \cite{Mathur2016}. However, these testbeds are either very costly to implement and maintain, making them unaffordable for most researchers, or they use commercial hardware and software components that are often treated as black boxes with little ability to model their inner working. The testbed we used for CPS security design, briefly presented in this paper, uses open hardware and software components, yet implements industrial standard communication protocols, e.g., Modbus/TCP.

Recently many CSTR-based testbeds were used to study cyber attacks on chemical processes.
The study in \cite{Guan2018} addressed the problem of joint distributed attack detection and secure estimation of the system states for a networked CPS over a wireless sensor network (WSN).  A malicious adversary simultaneously launches a False Data Injection (FDI) attack at the physical system layer to intentionally modify the system’s state and jamming attacks at the cyber layer to block the wireless transmission channels between sensors and remote estimators. The discretized and linearized state-space model of the CSTR near the operating point is used to show the effectiveness of the approach. 
To improve the overall resilience of CPS, the authors in \cite{januario2017resilience} propose a framework based on a distributed middleware that integrates a multiagent topology. The proposed framework is evaluated using a CPS simulator composed of a CSTR benchmark system model, a Wireless Sensor and Actuator Networks (WSAN) and additional remote devices, including a remote controller, a server where the model of the plant is running and an HMI. Jamming and node loss attacks are carried out as experiments for assessing the proposed attack framework. The work in \cite{durand2018nonlinear} develops a nonlinear system framework for understanding cyberattack-resilience of CSTR process using three control designs, where the focus is on data injection attacks on sensor measurements to impact the process safety. The study in \cite{Wu2018} integrates a neural network (NN)-based detection method and a Lyapunov-based model predictive controller for a class of nonlinear systems. A CSTR with constant volume is used to illustrate the application of the proposed NN-based detection method to handle cyber-attacks. A methodology for detecting abnormal events in aging Industrial Internet of Things (IIoT) has been developed in \cite{Genge2019}. The authors developed an efficient anomaly detection methodology that uses the correlation among process variables in order to detect stealthy cyber-attacks, and presented extensive experimental results on a CSTR model. An approach has been proposed in \cite{paoletti2019synthesizing} to either certify that a given control system is safe under possible cyber-attacks on the measured data used for feedback and/or the commanded control signals, or alternatively synthesize a particular spoofing attack that corrupts the signals to make the closed-loop system unsafe. A two state CSTR has been used as testbed for implementing the approach. The research work in \cite{zhang2018integration} employed a CSTR testbed to elucidate the dynamic interaction between feedback control and safety systems in the context of both model-based and classical control systems. To this aim, the interaction of a Model Predictive Control (MPC) system with a safety system is studied in the context of the Methyl isocyanate (MIC) hydrolysis reaction in a CSTR to avoid thermal runaway. The authors develop a specific action for the MPC to take when the safety system is activated due to significant feed disturbances that lead to thermal runaway conditions. \newline
Most of the reviewed testbeds mainly use control-theoretic approaches for the detection of cyber-attacks on chemical processes using state estimation techniques. The approaches in these papers are largely based on theoretical mathematical analysis and can get intractable for realistic CPS. Unlike the work reported in this paper, none of the studies considered the integration of the physical and cyber aspects of the CPS for risk assessment and countermeasures design. Existing work mainly considers the model-based approach for the physical system only, disregarding the cyber system. In addition, the work in this paper uses a standard CPS architecture that resembles industrial installations with industry-standard hardware and software.

In this paper, we develop a model-based approach for CPS security risk assessment. Rather than abstract formulations, we experiment with the design process on an industrial CPS testbed. The key objective of the experiments is to verify the design approach and to gain insights that could be used by the research community for future refinements. The work presented in this paper differentiates itself from existing literature in three main aspects: \begin{inparaenum}[(1)] \item it follows a model-based approach to couple CPS safety and security throughout the whole lifecycle, \item it proposes a formal approach to identify system hazards from a given physical model, rather than relying on expert opinion, and \item it implements the whole risk assessment lifecylce from asset identification to mitigation design, via a case study on an actual testbed, which provides both a clear implementation methodology as well as an insight for further research work\end{inparaenum}.

\section{CPS Testbed Architecture} \label{sec:CPS Description}
\begin{figure}[t]
	\centering
	\includegraphics[scale=0.4, trim = {0.25cm 0cm 0cm 0.1cm}, clip]{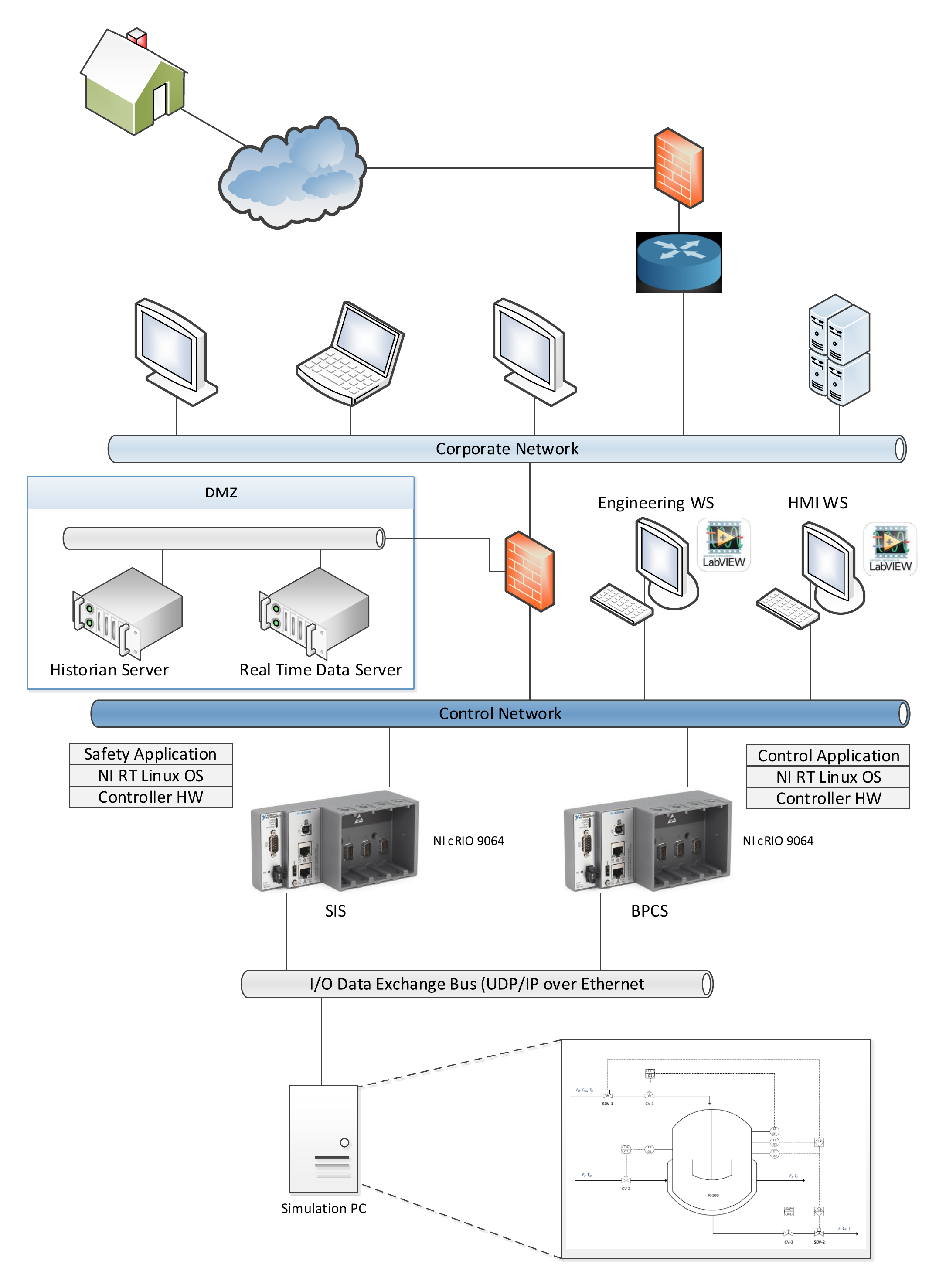}
	\caption{CPS Testbed Architecture}
	\label{fig:CPS-Arch}
\end{figure}

\begin{figure}[tb!]
	\centering
	\includegraphics[scale = 0.3, trim = {0cm 0cm 0cm 1cm}, clip, angle=-90]{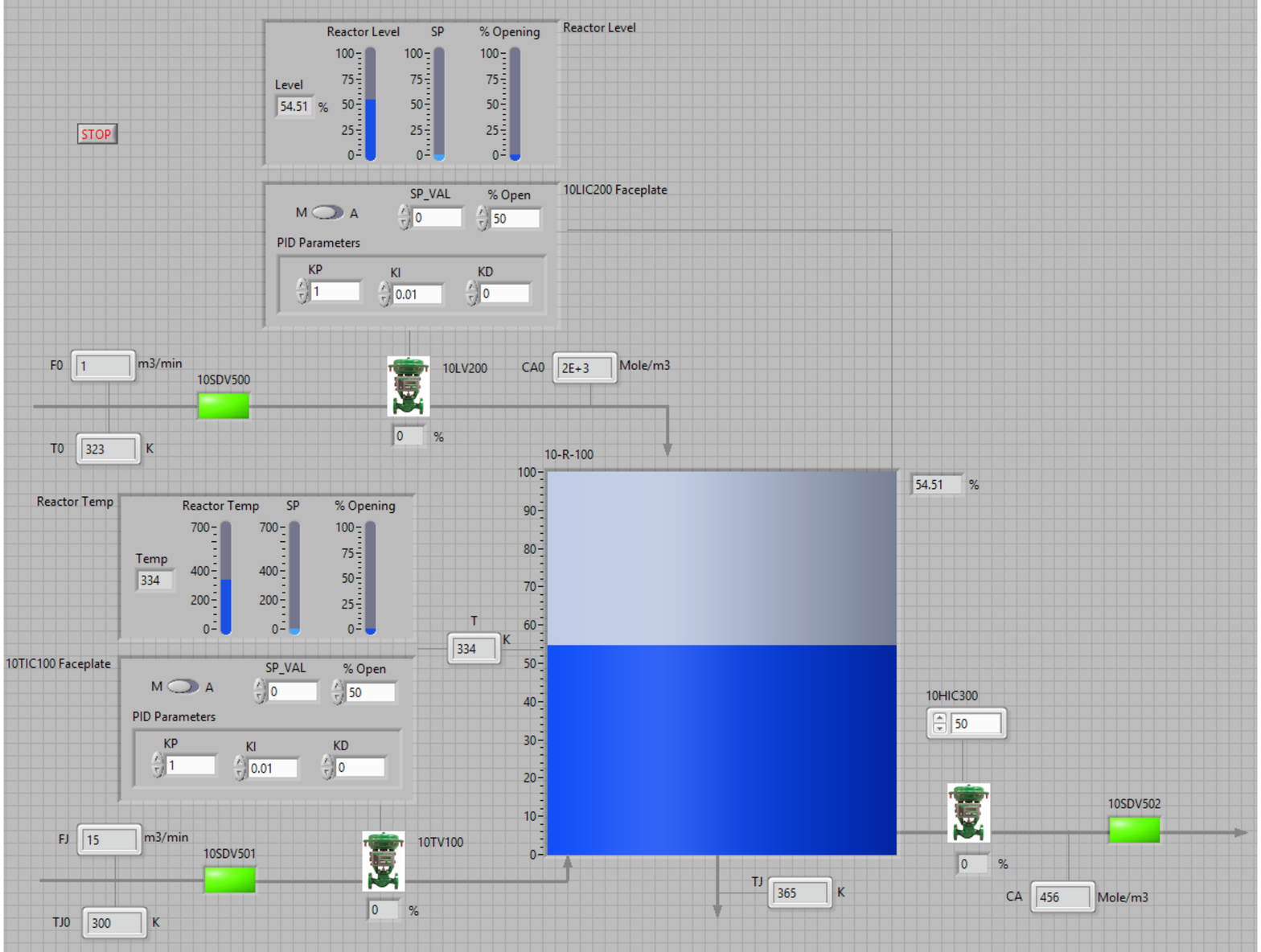}
	\caption{HMI for the Reactor Process}
	\label{fig:HMI-Reactor}
\end{figure}

Figure \ref{fig:CPS-Arch} illustrates the overall CPS architecture used in this paper. The physical system is an exo-thermal Continuous Stirred Tank Reactor (CSTR) simulated in real-time. Simulation data is exchanged over an I/O physical data bus to both the Basic Process Control System (BPCS) and the Safety Instrumented System (SIS). The BPCS implements the process control functions while the SIS implements the process shutdown logic functions, and both run RT Linux OS. The monitoring workstation implements the Human Machine Interface (HMI) for the operator to monitor and control the plant. The HMI has access control mechanisms that allow different levels of control based on the user role (typically operator, supervisor, and engineer). Figure \ref{fig:HMI-Reactor} shows the implemented HMI. A control network interconnects BPCS, SIS, Human Machine Interface (HMI) workstation, and the Engineering workstation. A firewall is configured to isolate the control network from the corporate network. A DeMilitarized Zone (DMZ) is formed to host the historian and real time data servers, following NIST proposed architecture \cite{StoufferKeithJoeFalco2015}. The RT Server runs the open source mysql database to store plant data for later retrieval by the HMI application and corporate financial analysis applications.

\subsection{Physical System}   \label{sec:physical-system}
We choose the Continuous Stirred Tank Reactor (CSTR) as the physical system for the testbed. The CSTR is an essential equipment in process plants where new products are formed from raw inlet reactants. The CSTR process possesses several features that make it a good choice for CPS security studies. First, the process variables to be controlled are closely-coupled, hence any change in one process variable will impact other variables as well and manifest itself in the overall process behavior. Second, the process has a number of potential safety hazard scenarios that may be produced by a cyber attack. Finally, mitigation layers for a number of safety hazards rely mainly on the control and safety systems, which are cyber systems that could be compromised by a cyber attack.

\subsubsection{Process Description}
We consider an irreversible exothermic CSTR process, with a first order reaction in the reactant $A$ to produce product $B$ with rate $k$ and a heat of reaction $\lambda$.
\begin{center}
\ce{A ->[$k$] B}
\end{center}

\begin{figure}[t]
\centering
\includegraphics[scale = 0.55, trim = {1cm 2 2.5cm 0}, clip]{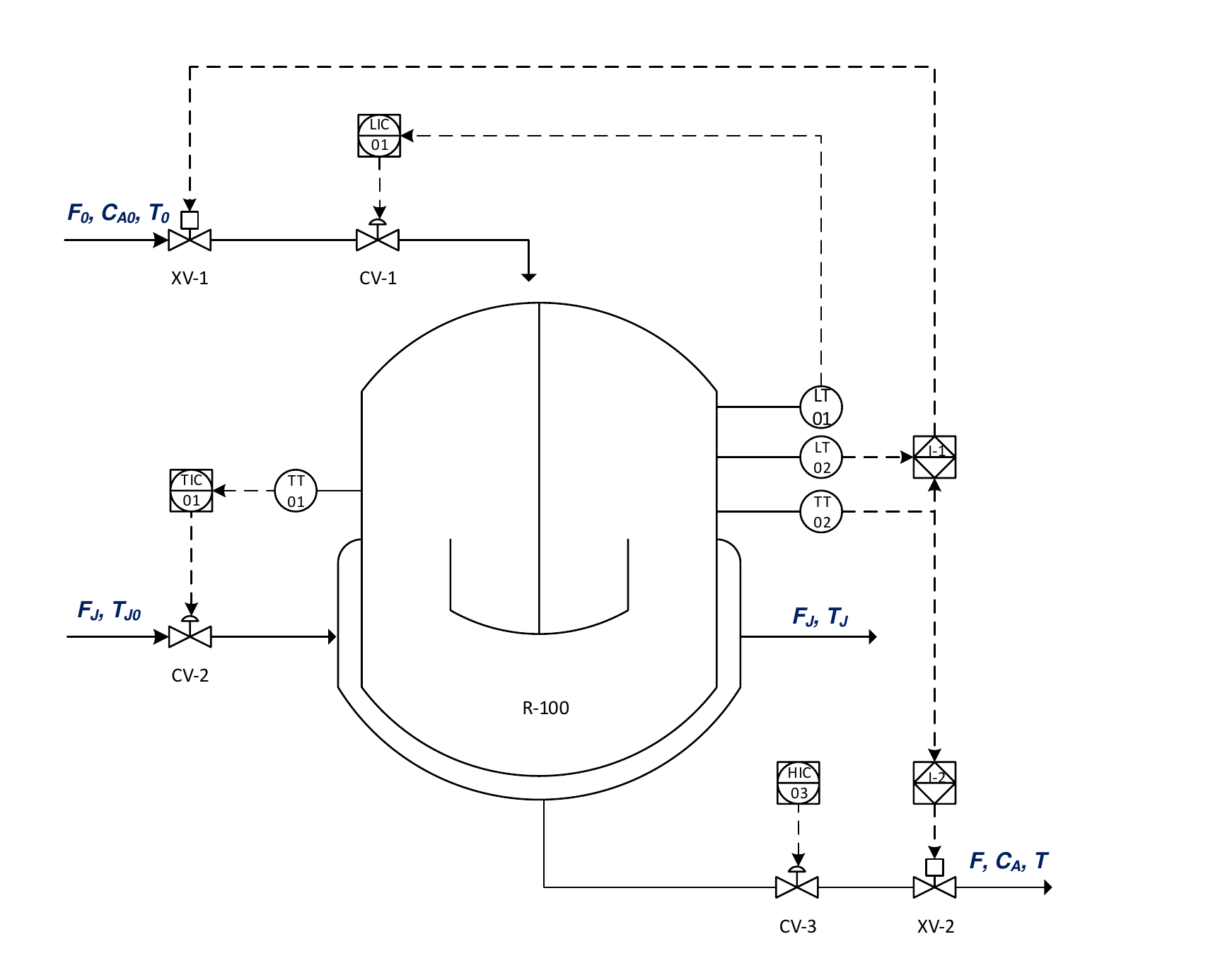}
\caption{Reactor P\&ID}
\label{fig:Reactor P&ID}
\end{figure}

Figure \ref{fig:Reactor P&ID} shows the Piping \& Instrumentation Diagram (P\&ID) for the reactor. The reactor vessel has an inlet stream carrying the reactant A, an outlet stream carrying the product B, and a cooling stream carrying the cooling fluid into the surrounding jacket to absorb the heat of the exothermic reaction. Reactant $A$ enters the reactor with concentration $C_{A_0}$, temperature $T_0$ and volumetric flow rate $F_0$. A first order reaction takes place where a mole fraction of reactant $A$ is consumed to produce product $B$. The outlet stream contains both reactant $A$ and product $B$, with reactant $A$ concentration $C_A$, outlet temperature $T$, and flow rate $F$. The outlet temperature $T$ is the same as the reactor temperature. The coolant fluid flows into the reactor jacket with temperature $T_{J_0}$ and flow rate $F_{J_0}$, and leaves the jacket with temperature $T_J$. The total coolant volume in the jacket is designated by $V_J$. The detailed mathematical model is developed in \cite{Tantawy2019CICN}.

\subsubsection{Process Measurement \& Control} \label{subsec:Process_Control}
The reactor temperature could be controlled by adjusting the coolant fluid inlet flow. A single variable control loop is used to regulate the reactor temperature. The control loop is composed of a temperature sensor TT-01, Temperature Controller TIC-01, and Control Valve CV-2. PID control algorithm is used for the controller as it still represents 98\% of all feedback control at over fifty thousand manufacturing facilities around the world \cite{miller2000web}, hence it is the de facto standard in the process control industry. Similarly, the reactor liquid level could be controlled by adjusting the inlet flow rate, using a control loop composed of a level sensor LT-01, Level Controller LIC-01, and Control Valve CV-1. Figure \ref{fig:Reactor P&ID} shows the two control loops using ISA standard symbols \cite{ISA51}.

\subsubsection{Process Safety Shutdown}    \label{subsec:safety-shutdown}
Process hazards are usually identified through a systematic risk assessment process involving hazard studies, e.g., HAZard and OPerability (HAZOP) study. High reactor level will lead to reactor overflow. The hazard of the overflow will depend on several factors, including the toxicity of the reactants, the operating temperature of the reactor, and the occupancy level of operators around the reactor area. Similarly, high reactor temperature will lead to exceeding the reactor design temperature and possible reactor damage. Table \ref{table:HAZOP} summarizes these two key hazards for the process and the possible initiating events.

\begin{table}[tb]
\begin{center}
    \setlength{\tabcolsep}{5pt}
    \begin{tabular}{ |p{1.5cm}|p{2cm}|p{2cm}|p{1.5cm}| } 
	\hline
	\textbf{Hazard} & \textbf{Initiating Event (Cause)} & \textbf{Consequences} & \textbf{Safeguards} (IPL) \\ 
	\hline
	High Level (Reactor overflow) & BPCS failure OR Outlet control valve fully closed OR Inlet valve stuck fully open & 2 or more fatalities (safety), Product loss (financial), Environmental contamination (environment) & Reactor dike (Mitigation) \\
	\hline
	High Temperature (Reactor Meltdown/explosion) &  Coolant inlet control valve fully (partially) closed OR Inlet valve stuck fully open & 10 or more fatalities (safety), Product loss (financial), Environmental contamination (environment) & None \\
	\hline
    \end{tabular}
    \caption{Partial HAZOP sheet for the reactor process}
\label{table:HAZOP}
\end{center}
\end{table}

For the Reactor overflow hazard, inlet stream has to be closed. Following IEC 61511 international standard, both the Safety Instrumented Function (SIF) and the control function have to be independent \cite{IEC61511}. Therefore, the SIF will be composed of an independent level sensor LT-02, a logic solver implementing an interlock function I-1, and a feed inlet shutdown valve XV-1. Upon detecting a high reactor level, the inlet feed will be stopped via the independent shutdown valve.

For the reactor high temperature hazard, the inlet stream has to be closed. Therefore, the SIF will be composed of an independent temperature sensor TT-02, a logic solver implementing an interlock function I-2, and the inlet stream shutdown valve XV-1. In addition, since it is preferred to keep the inventory, a shutdown valve XV-2 is added to the outlet stream. Therefore, upon detecting a high reactor temperature, both the inlet and outlet stream shutdown valves will be closed. These two safety functions are illustrated in Figure \ref{fig:Reactor P&ID}.

\subsection{Cyber System Architecture}  \label{sec:cyber-system}
Understanding the cyber system architecture is essential for CPS security design. Using the cyber architecture, combined with in-depth knowledge about the physical process dynamics and its associated hazards, a model-based approach could be followed to design an optimal security system for the given CPS according to predetermined optimality criteria, such as the minimization of financial loss or the probability of occurrence of a hazardous event.

\begin{figure}[tb]
	\centering
	\includegraphics[scale = 0.65, trim = {0.5cm 0.5cm 0cm 0cm}, clip, angle=0]{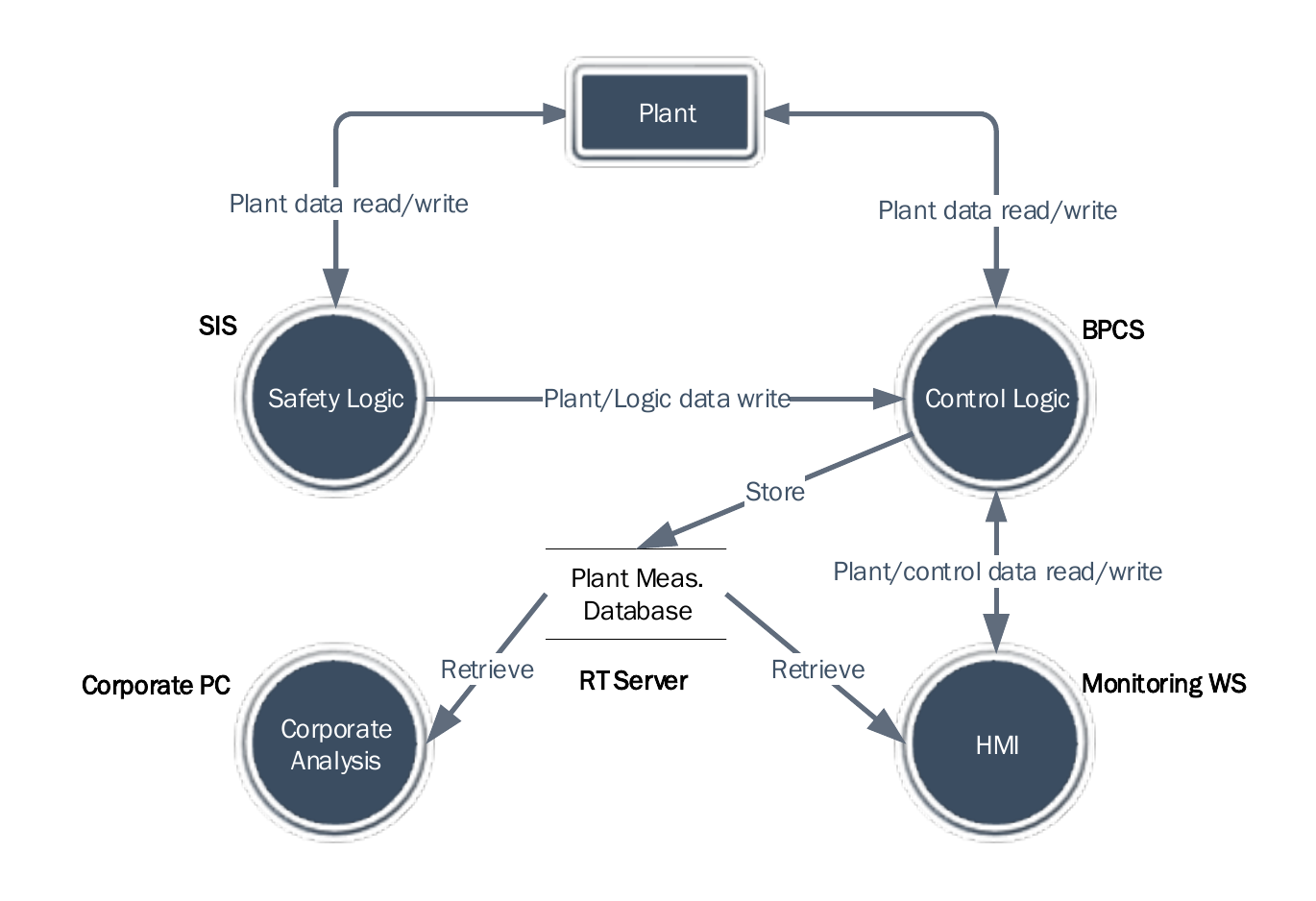}
	\caption{CPS Testbed Dataflow diagram}
	\label{fig:TESTBED-SW-Arch}
\end{figure}

Figure \ref{fig:TESTBED-SW-Arch} shows the data flow diagram for the testbed. There is a one-way data communication from SIS to BPCS, and HMI does not have direct communication to the SIS as per industrial practice. The HMI has direct communication with BPCS for monitoring and supervisory control actions. Plant data is stored in RT server database, for later on-demand retrieval by the HMI application running on the monitoring workstation and the corporate analysis application running on corporate PCs. As the RT Server has multiple communication links, it will play a key role in attacker penetration from the corporate LAN to the control LAN as detailed in the paper.

\section{Identification of Cyber-Related Hazards}	\label{sec:hazard-id}
\subsection{Hybrid Automaton Formulation}
CPS security attacks distinguish themselves from generic IT system attacks in that the main attacker's goal is to cause damage to the end physical system. Although CPS attacks may have the objective of stealing confidential information that may impact company business, such attacks could be launched on the business LAN level, and does not require the extra effort to penetrate down to the control network. Accordingly, the first natural step to analyze CPS security threats is to identify cyber attacks that are hazard-related. Given the complete process design with control and safety systems, as represented by the P\&ID in Figure \ref{fig:Reactor P&ID}, it is possible to construct the hybrid system automaton and utilize it to identify the process hazards and the associated cyber-controlled components that \emph{may} be manipulated by an attacker to cause the process hazard.

Hybrid system automaton is a formal model that describes hybrid dynamical systems that have both discrete and continuous dynamics \cite{raskin2005introduction}. A simplified mathematical model for the hybrid automaton is represented as a collection $\mathcal{H}=(Q,E,\Sigma,X,\text{Init,Inv,Flow,Jump})$, where $Q$ is a finite set of modes, $E$ is a finite set of event names, $\Sigma: Q \times E \rightarrow Q$ is a transition function representing the discrete changes, $X$ is a set of real-valued variables, Init, Inv, and Flow are functions that define the initial values, constrains, and evolution of the state variables $X$ for each mode, respectively, and finally $\text{Jump}$ assigns to each labelled edge a guard condition.

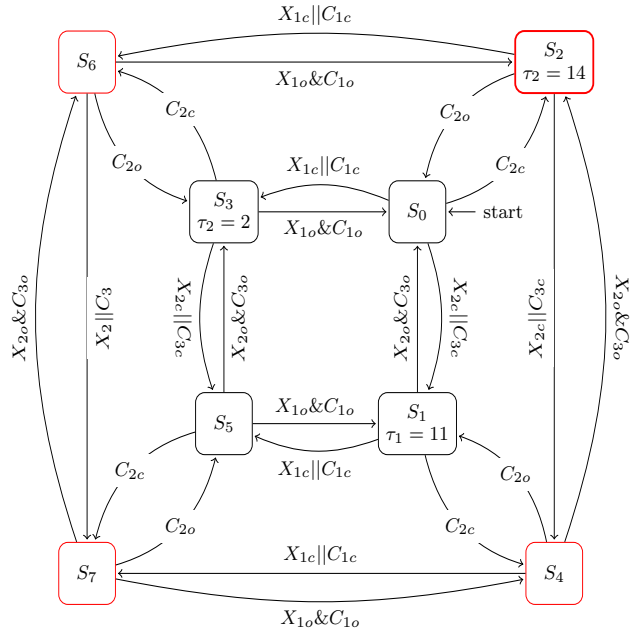
\begin{figure}[tb]
    \centering
    \resizebox{8.5cm}{8.5cm}{
    \begin{tikzpicture}[block/.style={minimum width={width("$\tau_2=14$")+2pt}},shorten >=1pt,node distance=3cm,on grid,auto] 
        \tikzstyle{every node}=[font=\footnotesize]
        \tikzset{every state/.append style={rectangle, rounded corners}}
        \node[state,initial right] (s_0)   {$S_0$};
        \node[state,align=center] (s_1) [below =of s_0] {$S_1$ \\ \footnotesize{$\tau_1 = 11$}};
        \node[state,align=center,draw=red,thick] (s_2) [above right =of s_0] {$S_2$ \\ \footnotesize{$\tau_2 = 14$}};
        \node[state,align=center] (s_3) [left=of s_0] {$S_3$ \\ \footnotesize{$\tau_2 = 2$}};
        \node[state,align=center,draw=red] (s_4) [below right=of s_1] {$S_4$};
        \node[state,align=center] (s_5) [left=of s_1] {$S_5$};
        \node[state,align=center,draw=red] (s_6) [above left =of s_3] {$S_6$};
        \node[state,align=center,draw=red] (s_7) [below left=of s_5] {$S_7$};
        
        \path[->] 
        (s_0) edge [bend left = 20]  node [anchor=center,above,rotate=-90] {$X_{2c} || C_{3c}$} (s_1)
        (s_1) edge [above] node [rotate=90]{$X_{2o} \& C_{3o}$} (s_0)
        (s_0) edge [bend right=30]  node [fill=white,anchor=center]{$C_{2c}$} (s_2)
        (s_2) edge [bend right=30]  node [fill=white,anchor=center]{$C_{2o}$} (s_0)
        (s_0) edge [bend right=20, above]  node {$X_{1c} || C_{1c}$} (s_3)
        (s_3) edge [below] node {$X_{1o} \& C_{1o}$}(s_0)
        (s_1) edge [bend right=30]  node [fill=white,anchor=center] {$C_{2c}$} (s_4)
        (s_4) edge [bend right=30]  node [fill=white,anchor=center] {$C_{2o}$} (s_1)
        (s_2) edge node [fill=white,anchor=center,above,rotate=90] {$X_{2c} || C_{3c}$} (s_4)
        (s_4) edge [bend right=20]  node [anchor=center,above,rotate=-90] {$X_{2o} \& C_{3o}$} (s_2)
        (s_1) edge [bend left=20,below]  node {$X_{1c} || C_{1c}$} (s_5)
        (s_5) edge [above] node {$X_{1o} \& C_{1o}$} (s_1)
        (s_3) edge [bend right=20]  node [anchor=center,below,rotate=-90] {$X_{2c} || C_{3c}$} (s_5)
        (s_5) edge [below] node [rotate=90]{$X_{2o} \& C_{3o}$} (s_3)
        (s_2) edge [bend right=10,above]  node {$X_{1c} || C_{1c}$} (s_6)
        (s_6) edge [below]  node {$X_{1o} \& C_{1o}$} (s_2)
        (s_3) edge [bend right=30]  node [fill=white,anchor=center] {$C_{2c}$} (s_6)
        (s_6) edge [bend right=30]  node [fill=white,anchor=center] {$C_{2o}$} (s_3)
        (s_5) edge [bend right=30]  node [fill=white,anchor=center] {$C_{2c}$} (s_7)
        (s_7) edge [bend right=30]  node [fill=white,anchor=center] {$C_{2o}$} (s_5)
        (s_4) edge [above]  node {$X_{1c} || C_{1c}$} (s_7)
        (s_7) edge [below, bend right = 10]  node {$X_{1o} \& C_{1o}$} (s_4)
        (s_6) edge [left]  node [fill=white,anchor=center,below,rotate=90] {$X_2 || C_3$} (s_7)
        (s_7) edge [bend left=20]  node [anchor=center,above,rotate=90] {$X_{2o} \& C_{3o}$} (s_6);
	\end{tikzpicture}}
    \caption{Reactor hybrid automaton. For transition labels, $X$ refers to a shutdown valve and $C$ refers to a control valve. Subscript 'o' refers to valve status 'open', while 'c' refers to valve status 'closed'. It is assumed that the associated Boolean variable is high when the valve is closed. Reactant stream: $X_1 || C_1$, Coolant stream: $C_2$, Outlet stream: $X_2 || C_3$.}
    \label{fig:Hybrid Automaton}
\end{figure}

According to the hybrid automaton definition, we can describe the CSTR process by the hybrid automaton in Figure \ref{fig:Hybrid Automaton}. The set of modes $Q = \{S_0,S_1,\ldots,S_7\}$ is defined by the status of the three process streams, namely inlet stream $I_s$, outlet stream $O_s$, and cooling jacket stream $J_s$. The corresponding set of events are $E = \{I_s=0,I_s=1,O_s=0,O_s=1,J_s=0,J_s=1\}$, where a value of zero refers to a closed stream and a value of 1 refers to an open stream. The set of state variables $X = \{L,T,T_j,C_A\}$ describes the reactor level, temperature, coolant temperature, and inlet product mole concentration, respectively. The initial state is $S_0$, designating normal reactor operation when all streams are open. The state constraints and evolution are expressed according to the reactor dynamics and omitted for brevity. Finally, the Jump describing the guard conditions for each edge is described in terms of respective valve positions.

In addition to the hybrid system automaton, the system is usually constrained to run inside an operating envelope, $\mathcal{E} \subset \mathbb{R}^n$, where $n$ is the number of state variables, i.e., $n=|X|$. A system hazard occurs when one or more of the state variables are outside the operating envelope, i.e., $X \notin \mathcal{E}$. This criterion was applied to the CSTR system via simulation to identify the hybrid automaton hazardous states. Four hazardous states were identified, $S_2,S_4,S_6$, and $S_7$. In state $S_2$, the coolant stream is closed. The reactor and coolant fluid temperatures exceed the design limit of 550 K after 14 minutes and 23 minutes, respectively, until reaching the thermal equilibrium point $T = T_J = 582$ K via heat exchange after 60 minutes. We define the process hazard time by the minimum time taken by the process to exceed the design limit, therefore $\tau_2 = 14$ min. In state $S_4$, the outlet and coolant streams are closed. If the coolant steam is closed first, then the system will transition through state $S_2$ and the reactor will stabilize at the unsafe temperature $T = T_J = 582$. State $S_6$, where the inlet and coolant streams are closed, and state $S_7$, where the reactor is completely isolated, are similar to state $S_4$. In summary, the reactor will be subject to a high temperature beyond the design limit in all hazardous states, potentially leading to reactor runaway.
Hazardous states are designated in Figure \ref{fig:Hybrid Automaton} by the red border lines.

\begin{figure}
    \centering
    \begin{forest}
        for tree={circle,draw, l sep=18pt, s sep=37}
        [$s_0$ 
            [$s_1$, edge label={node[midway,above,xshift=-0.5cm]{$X_{2c} || C_{3c}$}}
                [$s_4$, red, edge label={node[midway,above,xshift=-0.5cm]{$C_{2c} $}}]
                [$s_5$, edge label={node[midway,right]{$X_{1c} || C_{1c}$}}
                    [$s_7$, red, edge label={node[midway,right,xshift=-0.75cm]{$C_{2c} $}}]
                ]
            ]
            [$s_2$, red, edge label={node[midway,right]{$C_{2c}$}}]
            [$s_3$, edge label={node[midway,above,xshift=0.5cm]{$X_{1c} || C_{1c}$}}
                [$s_6$, red, edge label={node[midway,right]{$C_{2c} $}}]
            ]
        ]
    \end{forest}
    \caption{Hazard execution tree for the CSTR process}
    \label{fig:hazard_tree}
\end{figure}
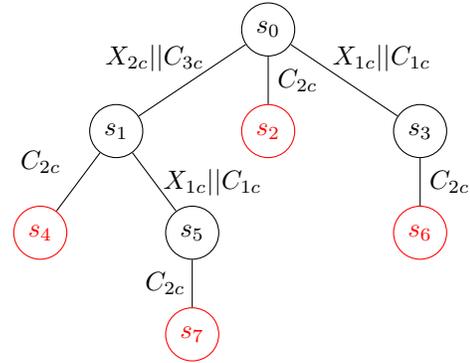

Given the hybrid automaton $\mathcal{H}$ of the system with the designated hazardous states, we can extract the hazard initiating events. Algorithm \ref{alg:auto-hazop} presents a formal method to obtain the attack scenarios that potentially cause a process hazard. The input to the algorithm is the hybrid system automaton, designated hazardous states, and the initial system state $S_0$. The algorithm returns the hazard execution tree, where the tree paths are the shortest paths from the initial state $S_0$ to each hazardous state. These traces are later linked to cyber components that manipulate field devices to cause the process hazard. The algorithm can use any variant of the shortest path algorithm \cite{ThomasH.Cormen2009}. Algorithm \ref{alg:auto-hazop} was applied to the CSTR hybrid automaton, and Figure \ref{fig:hazard_tree} shows the hazard execution tree output. Hazard states will always be leaf nodes in the hazard tree. Example hazard execution traces are: $s_o \xrightarrow{C_{2c}} s_2$ and $s_0 \xrightarrow{X_{2c}||C_{3c}} s_1 \xrightarrow{C_{2c}} s_4$. 

\begin{algorithm}[tb]
\SetAlgoLined
\SetKwInOut{Input}{input}
\SetKwInOut{Output}{output}
\Input{$\mathcal{H},S_0, \mathbf{S},$ (Automaton, initial state, hazardous states)}
\Output{$M$ (Adjacency matrix for the hazard execution tree)}
\For{$S \in \mathbf{S}$}
{
	ShortestPath($S_0,S,M$) \;
}
\Return $M$;
\caption{Generation of hazard execution tree}
\label{alg:auto-hazop}
\end{algorithm}

It should be highlighted that for large systems with a large number of inputs, the number of states of the hybrid automaton may be prohibitively large. One approach to solve this problem is the decomposition of the large system into smaller components, and apply algorithm \ref{alg:auto-hazop} to each individual component. In fact, this is the industrial practice when conducting manual hazard analysis by experts, e.g., HAZard and OPerability (HAZOP) study \cite{dunjo2010hazard}. In HAZOP, the system is divided into nodes and each node is studied separately. This approach works well as long as the coupling between system components is captured in the analysis of each individual component.

Figure \ref{fig:hazard_tree} shows how a process hazard is generated in terms of process actions, such as opening or closing a valve, but it does not show how these actions could be taken by an attacker. To do so, the cyber architecture of the CPS is required. From the cyber architecture, the cyber components that are linked to process actions could be identified along with their associated vulnerabilities that could be exploited to launch the cyber attacks. Once vulnerabilities are identified, attack trees could be used for threat modeling \cite{mauw2005foundations}.

\begin{figure}
    \centering
    \includegraphics[scale=0.8]{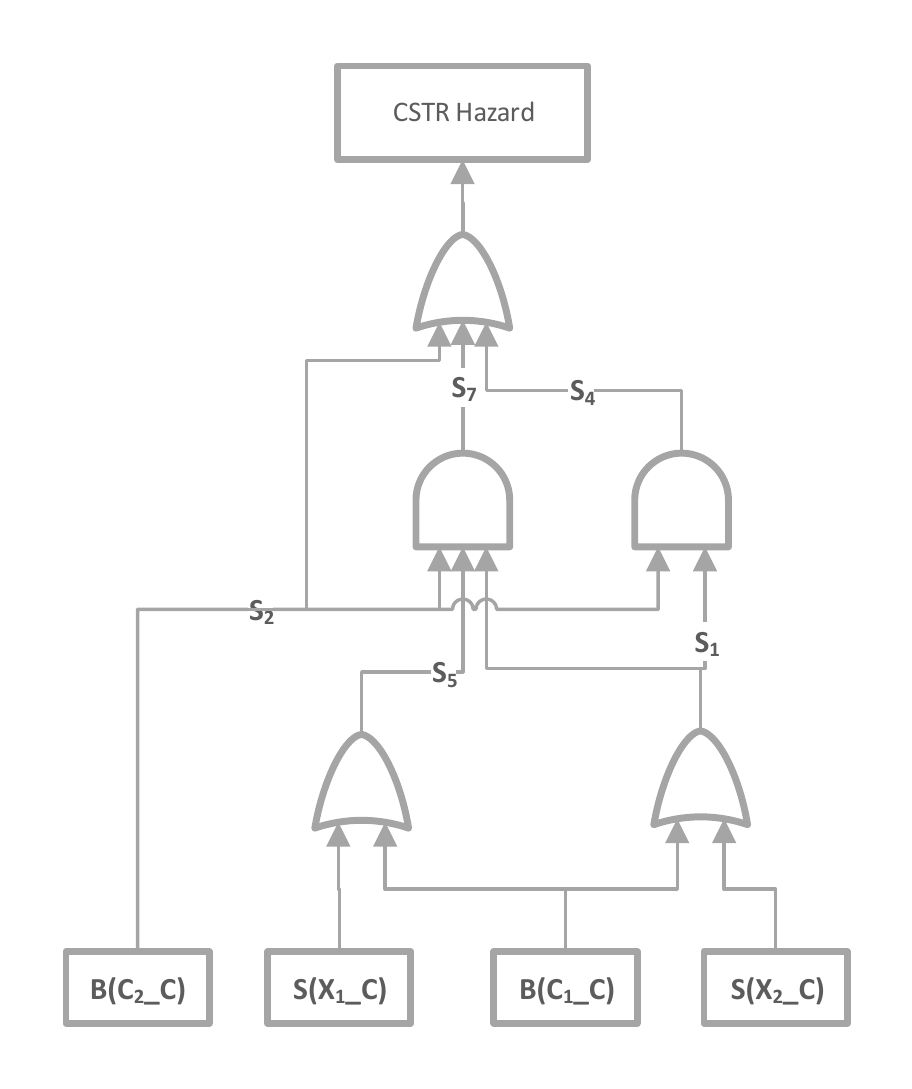}
    \caption{Partial attack tree for the CSTR hazard tree in figure \ref{fig:hazard_tree}}
    \label{fig:attack_tree_hazard}
\end{figure}

In order to convert the hazard tree into an attack tree, we need a mapping from process actions to cyber component actions. From Figures \ref{fig:CPS-Arch} and \ref{fig:Reactor P&ID}, the control valves are connected to the BPCS while the shutdown valves are connected to the SIS. We designate a relevant valve action using the terminology <System>(<Valve\_tag>\_<Action>), with system values being $B$ for BPCS and $S$ for SIS, and action values being 'C' for Close and 'O' for Open. Therefore, $B(C_2\_C)$ refers to closing valve $C_2$ via BPCS relevant vulnerability exploitation. Figure \ref{fig:attack_tree_hazard} shows a partial \emph{abstract} attack tree generated from the hazard tree in Figure \ref{fig:hazard_tree} using the introduced terminology.

From the attack tree, it is evident that $B(C_2\_C)$, i.e., closure of valve $C_2$ via BPCS vulnerability exploitation, is sufficient to cause a reactor hazard. Therefore, in the next section we focus on the compromise of the BPCS to cause a process hazard. It should be highlighted that in general, it may be required to take multiple actions by different cyber components to achieve a process hazard. The CSTR process is a special case in the sense that the closure of $C_2$ is necessary, and sufficient, to cause a process hazard.

\subsection{Hybrid Automaton Expressiveness}
The analysis shows that the hybrid system automaton is a rich diagram, where important information could be inferred. First, a hazard execution tree could be generated, where hazard states are identified and execution traces that lead to process hazards along with the associated actions could be extracted. Second, the time it takes the process to produce a hazard is labeled inside the hazardous state, enabling an accurate assessment of the required response time. Finally, different ways to take the process out of the hazardous state before the time to hazard elapses could be inferred from the outward transitions, noting that the system could return to the previous state by reversing the transition action (e.g., opening a valve that was closed), Therefore, it is possible to identify the physical countermeasures in order to protect the process in case of an attack.

\section{Design of Cyber Attacks} \label{sec:cyber-attacks}
Figure \ref{fig:attack_tree_hazard} illustrates the fact that valve CV-2 closure via BPCS is a necessary and sufficient cause for reactor hazard. Therefore, this section focuses on the design of cyber attacks for BPCS to cause the desired valve closure, either by integrity or DoS attacks.

From the CPS architecture, the attack entry point is the corporate PC. This could represent an insider attack, using a personal laptop or a corporate PC, or an outside attacker over the Internet who compromised a corporate PC. The mechanism by which the outside attacker could gain access to the corporate PC either locally or remotely via a remote session has been studied extensively in the IT Security literature and will not be treated in the paper. For further details about IT security vulnerabilities, refer to \cite{computerSecurityBook2018}. In the rest of the paper, we will assume the attacker has a full privileged access to a corporate PC, but not to the control network.

\begin{figure}[tb!]
	\centering
	\includegraphics[scale = 0.6, trim = {0.5cm 0.1cm 0.1cm 0.1cm}, clip, angle=0]{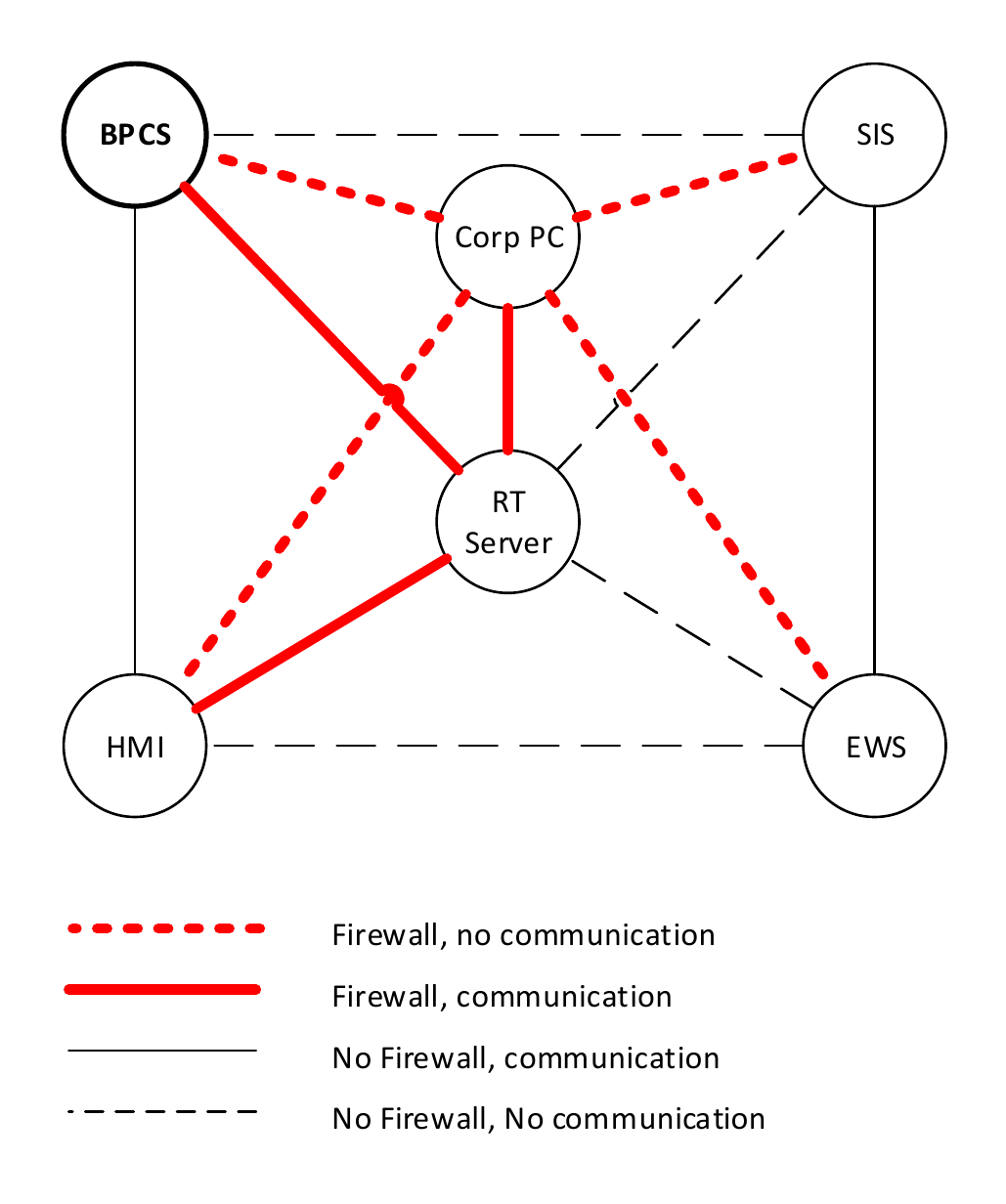}
	\caption{BPCS connectivity graph. Different line styles designate different connectivity patterns and hence the difficulty of attacks. Only one DMZ server is shown.}
	\label{fig:BPCS-Data-Flow}
\end{figure}

\noindent\textbf{Connectivity Graph:}
CV-2 is part of a temperature control loop that stabilizes the reactor temperature. The full control loop, including process I/O communication and the PID controller, is implemented on the BPCS. The compromise of CV-2 thus could be achieved by compromising the BPCS itself or any of its connected cyber components to act as a pivot. Figure \ref{fig:BPCS-Data-Flow} shows the connectivity graph for the BPCS, constructed from the CPS architecture and the data flow graph in Figures \ref{fig:CPS-Arch} and \ref{fig:TESTBED-SW-Arch}, respectively. The number of all possible paths between the corporate PC and the BPCS target grows exponentially with the graph size. However, given that most paths are very unlikely due to the inherent difficulty, e.g., the presence of a firewall or absence of any data communication between the connected nodes, we exclude paths that have a firewall and lack data communication between nodes. The probability of a cyber attack across these excluded paths will be insignificant and their impact on the risk assessment process could be neglected. It should be highlighted that for typical CPS networks, the number of nodes is small so the extraction of all possible paths is still feasible, although computationally expensive.

\begin{figure}[htb!]
	\centering
	\includegraphics[scale = 0.75, trim = {0.6cm 0.5cm 0.6cm 0.6cm}, clip, angle=0]{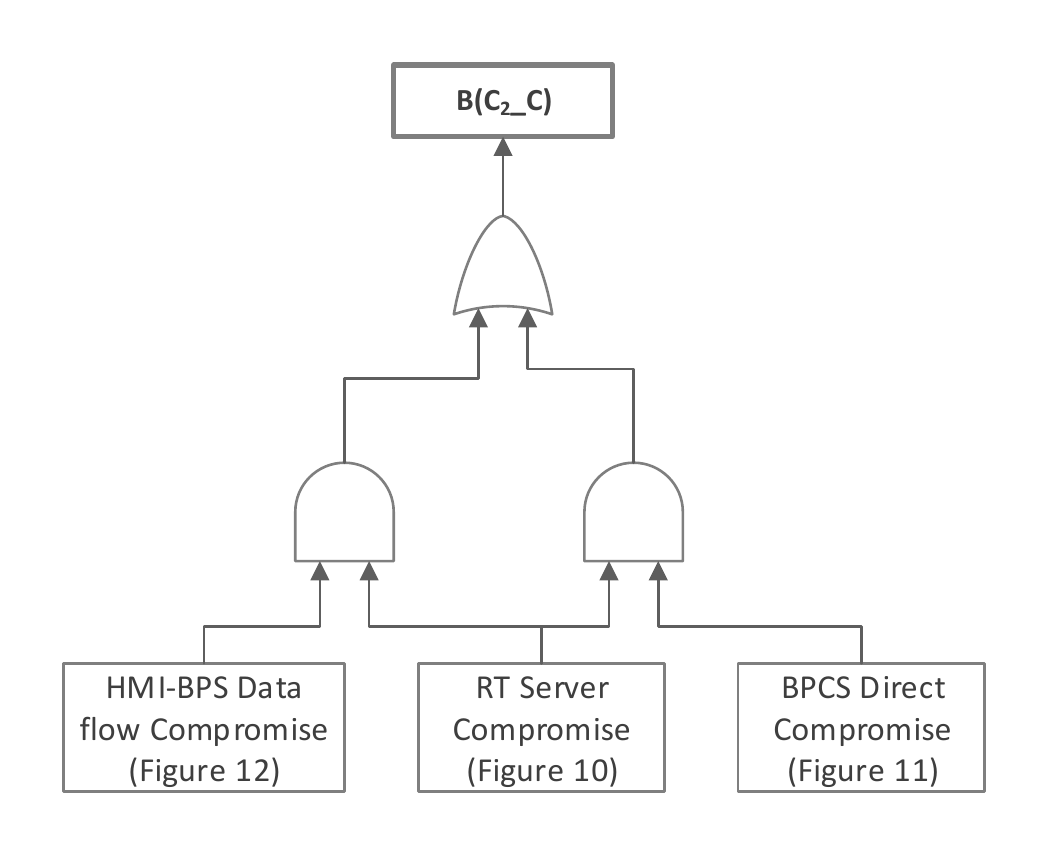}
	\caption{Reactor Runaway Abstract Attack Tree. Attack scenarios are sequential, starting from RT server compromise, e.g., RT server compromise followed by a BPCS compromise represents one attack scenario.}
	\label{fig:Reactor-Runaway-Attack-Tree-Abstract}
\end{figure}

Figure \ref{fig:Reactor-Runaway-Attack-Tree-Abstract} shows the attack tree that enumerates the likely paths to compromise the reactor process starting from the corporate PC. The root of the tree represents the cyber attack objective of closing valve $CV_2$, which is the leaf node in the abstract attack tree in Figure \ref{fig:attack_tree_hazard}. The basic events in the tree in Figure \ref{fig:Reactor-Runaway-Attack-Tree-Abstract} are modeled using detailed attack trees in this section.

In the following, vulnerability scanning tools were used to identify vulnerabilities and to design cyber attacks for the identified nodes in Figure \ref{fig:Reactor-Runaway-Attack-Tree-Abstract}, namely Real time server, BPCS, and HMI. Whether the vulnerabilities are exploitable is verified by penetration testing. Client-side attacks, such as phishing emails, are not considered for the BPCS, as it is an embedded unattended node (no direct HMI attached), and for the HMI, as it is not an industrial practice to run any internet application such as a browser or email client on the HMI. Table \ref{tab:VULN-PENTEST-CMD} in the appendix summarizes the tools and commands used.

\subsection{Design of RT Server Cyber Attacks}
\begin{figure}[b!]
	\centering
	\includegraphics[scale = 0.62, trim = {0.65cm 0.5cm 0.6cm 0.6cm}, clip, angle=0]{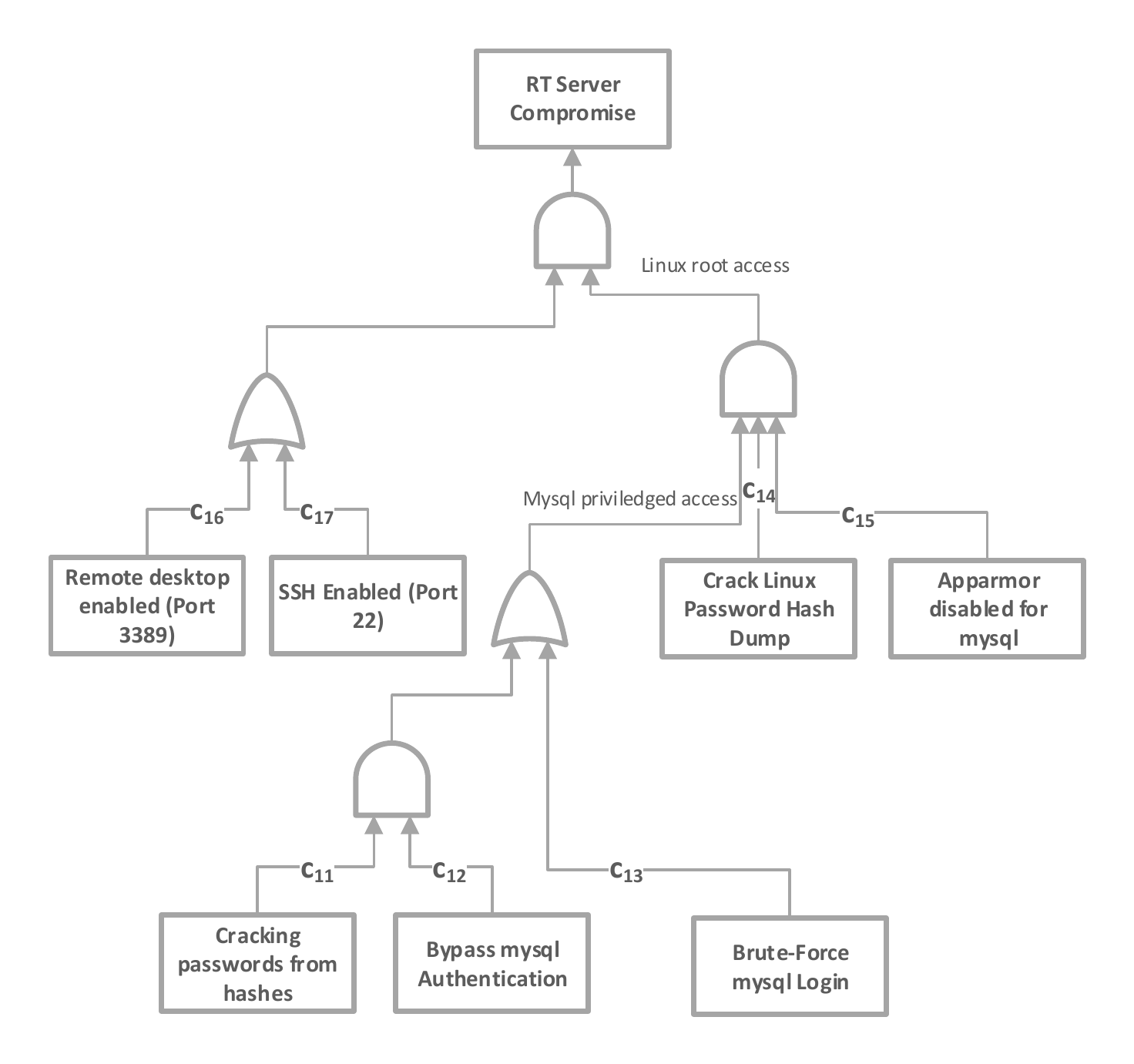}
	\caption{RT Server Attack Tree.}
	\label{fig:RTSRV-Attack-Tree}
\end{figure}

We start from the compromised corporate PC that has legitimate access to the Real Time (RT) server for data logging purposes. Using nmap port scanning command against RT server, we get the information reported in Table \ref{tab:RTSRV-VULN} for mysql server and ssh services. Mysql could be exploited in different ways including authentication bypass, password hash dump, and brute-force login, depending on the specific server settings. Figure \ref{fig:RTSRV-Attack-Tree} shows an attack tree to compromise the RT server by escalating the attacker privileges via mysql exploitation.

\begin{table}[tb]
	\begin{center}
    \setlength{\tabcolsep}{10pt}
    \begin{tabular}{ |P{0.5cm}|p{0.25cm}|P{0.7cm}|p{1.5cm}|p{2cm}|} 
			\hline
			\textbf{Port} & \textbf{State} & \textbf{Service} & \textbf{Version} & \textbf{Vulnerabilities} \\ 
			\hline
			3306/tcp & open & sql & 5.7.28, Protocol 10 & Authentication bypass, passwords dump, Brute force passwords \\ \hline
			22/tcp & open & ssh & OpenSSH 7.9p1 Ubuntu 10 & \\ \hline
		\end{tabular}
	\end{center}
	\caption{RT Server vulnerabilities - nmap scan}
	\label{tab:RTSRV-VULN}
\end{table}

From the attack tree, the probability of RT Server compromise could be approximated by:
\begin{align}
P[\texttt{Srv-Comp}]= c_{14}c_{15} (c_{16}+c_{17}) (c_{11}c_{12} + c_{13})
\label{eq:RTSRV-ATTACK}
\end{align}
Penetration testing would assisst in assigning probability measures to the success of each attack action.

\subsection{Design of BPCS Cyber Attacks}
Using \texttt{nmap} port scanning tool, Table \ref{tab:BPCS-VULN} lists the open ports and associated listening services and service version. In addition, \texttt{nmap} identified the target as NI cRIO 9049 running RT Linux OS. There are many other open ports reported that are used for vendor-specific services, which could be exploited with proper knowledge about vendor hardware and software. We limit our discussion to known services only to keep the treatment general.

\begin{figure*}[tb!]
	\centering
	\includegraphics[scale = 0.65, trim = {0.65cm 0.5cm 0.6cm 0.6cm}, clip, angle=0]{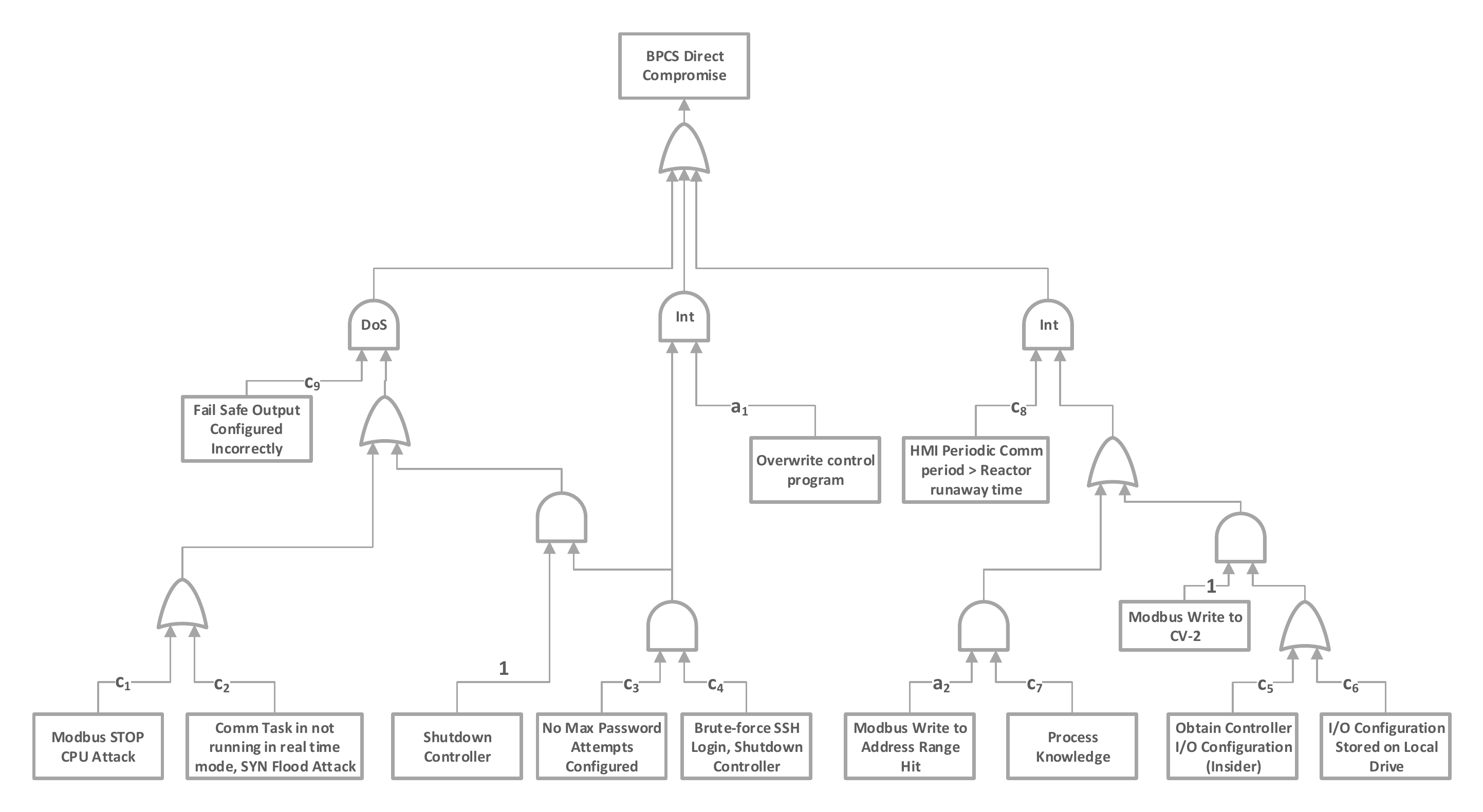}
	\caption{BPCS Attack Tree. Probabilities related to the corporation are designated by $c_i$ and to the attacker by $a_i$}
	\label{fig:BPCS-Attack-Tree}
\end{figure*}

According to open ports and services, Figure \ref{fig:BPCS-Attack-Tree} shows the BPCS attack tree. Port 502 listens to Modbus communication, so a Modbus malicious write attack, with and without a specific register address, is considered. Information about register addresses could be obtained from controller configuration data, either as a hard copy or stored on local controller drive. Lack of a specific register address for valve CV-2 will require writing to an address range, which may lead to a random valve configuration that won't cause a reactor hazard, e.g., closing all inlet and outlet valves. To increase the probability of success of such an attack, process knowledge is required. Further, Modbus write attack will not be successful unless no other Modbus master is writing to the same register. For the given testbed, HMI can write to valve registers after setting process controllers to manual mode. This requires that the reactor hazard time is less than the HMI-BPCS (a)periodic communication rate.

With SSH port open, another possible attack is Brute-force SSH login attempt to guess admin password and gain privileged access to the BPCS. This attack will succeed only if the SSH server is configured with no maximum number of password attempts. After gaining the required privileged access, BPCS controller could be shutdown to cause DoS attack, or the control problem could be overwritten to cause integrity attack, assuming the attacker has sufficient knowledge about the hardware as well as the necessary software tools.

Finally, A DoS attack could also be designed by exploiting Modbus STOP CPU function or by using SYN Flood attack against the Modbus or SSH open ports. This later attack will be successful only if the BPCS controller schedule is not a real time scheduler that will give priority to control tasks. It should be noted that DoS attack will not succeed in causing a reactor hazard if process outputs are configured as fail-safe. In such case, the controller will automatically write the pre-configured safe value to process valves in case of controller software crash.

\begin{table}[tb!]
	\begin{center}
    \setlength{\tabcolsep}{10pt}
    \begin{tabular}{ |P{0.4cm}|P{0.4cm}|P{0.4cm}|P{1.25cm}|p{1.5cm}|} 
			\hline
			\textbf{Port} & \textbf{State} & \textbf{Service} & \textbf{Version} & \textbf{Vulnerabilities} \\ 
			\hline
			22/tcp & open & ssh & OpenSSH 7.4 & Brute-force SSH login \\ \arrayrulecolor{gray}\hline
			502/tcp & open & mdps & Modbus/TCP & Modbus Write, Stop, MITM \\ \arrayrulecolor{black}\hline
		\end{tabular}
	\end{center}
	\caption{BPCS Vulnerabilities - nmap scan}
	\label{tab:BPCS-VULN}
\end{table}

From the BPCS attack tree, the probability of BPCS direct compromise could be approximated by:
\begin{align}
P[\texttt{BPCS-Comp}] = & c_9[c_3c_4 + (c_1 + c_2)] + a_1c_3c_4 + \nonumber \\
& c_8 [a_2c_7 + (c_5 + c_6)]
\label{eq:BPCS-ATTACK}
\end{align}

\subsection{Design of HMI-BPCS Data Flow Attacks}
In this attack vector, HMI has to be compromised first to mount the malicious data over the HMI-BPCS data flow stream. Table \ref{tab:HMI-VULN} shows the nmap scan results for the HMI. The machine is running Windows 10 and fully patched, therefore very few vulnerabilities were detected. Mysql service is running to communicate with the RT Server for real time data and trend display for the operator. Modbus master service is running to poll and write data to the BPCS controller. Remote desktop is turned on to grant different operators access to HMI machines for other plant zones.

\begin{figure}[tb!]
	\centering
	\includegraphics[scale = 0.6, trim = {0.6cm 0.5cm 0.6cm 0.6cm}, clip, angle=0]{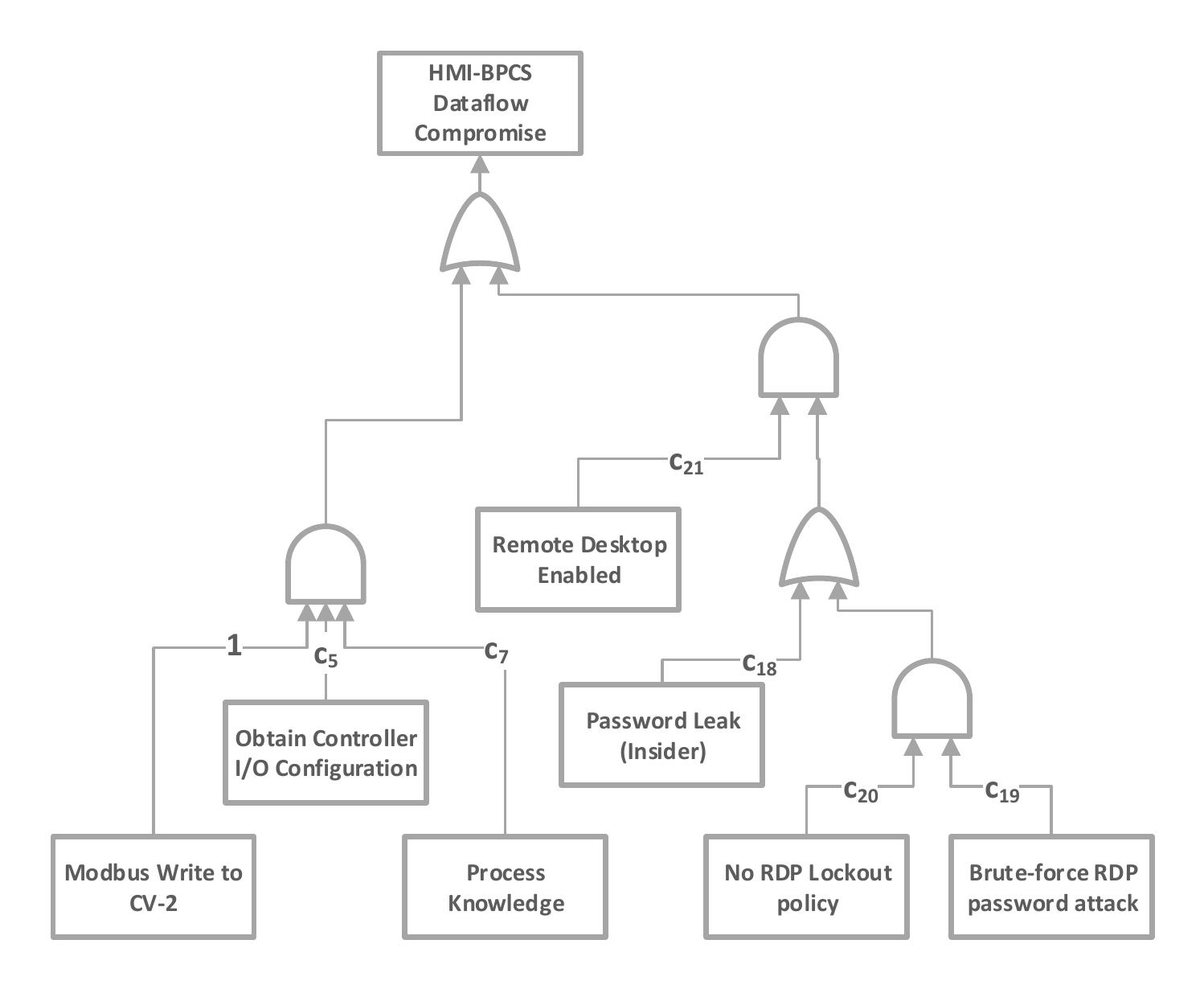}
	\caption{HMI-BPCS Dataflow Attack Tree}
	\label{fig:HMI-Attack-Tree}
\end{figure}

Figure \ref{fig:HMI-Attack-Tree} shows the HMI attack tree. Remote desktop is exploited using either brute-force password attack or via password leak. In addition, Modbus open port is exploited to write malicious values to relevant registers, provided process and I/O configuration knowledge are available. Since HMI has mysql client connection to RT Server, mysql client side attacks could be considered as well. However, given the low number of mysql client side vulnerabilities and the difficulty associated, we ignored mysql client side attacks in the penetration testing.

\begin{table}[tb!]
	\begin{center}
			\setlength{\tabcolsep}{10pt}
			\begin{tabular}{ |P{0.4cm}|P{0.4cm}|P{0.4cm}|p{1.25cm}|p{1.5cm}| } 
				\hline
				\textbf{Port} & \textbf{State} & \textbf{Service} & \textbf{Version} & \textbf{Vulnerabilities} \\ 
				\hline
				502/tcp & open & mdps & Modbus/TCP & Modbus write, stop \\ \hline
				3306/tcp & open & mysql & MySQL &  \\  \hline
				3389/tcp & open & rdp & & \\ \hline
				
			\end{tabular}
	\end{center}
	\caption{HMI Vulnerabilities - nmap scan}
	\label{tab:HMI-VULN}
\end{table}

From the HMI attack tree, the probability of HMI-BPCS dataflow compromise could be approximated by:
\begin{align}
P[\texttt{HMI-BPCS}] = & c_5 c_7 + c_{21}(c_{18}+c_{19}c_{20})
\label{eq:HMI-BPCS-Attack}
\end{align}

It should be highlighted that the presented attacks are not meant to be comprehensive. As an example, we did not consider Man In The Middle (MITM) attacks as the compromised RT server is behind a firewall, hence it cannot monitor the traffic or cause ARP poisoning unless one node on the control network is compromised. Also, we assume perfect security around the company premises, which does not allow for direct physical access to hardware. Finally, we assume the IT security policy does not allow the plug in of foreign USB devices to plant control equipment. Should any of these assumptions be violated, the attack trees have to be expanded to accommodate for these new threats. Moreover, the design of the cyber attacks does not include day zero attacks, which are unknown at design time and could only be detected online using anomaly detection techniques \cite{aleroud2012contextual}.

\section{Penetration Testing and Risk Assessment} \label{sec:pentest}
Penetration testing was carried out according to the attack trees designed in Section \ref{sec:cyber-attacks} to verify the identified vulnerabilities and support the risk assessment process by quantifying the probability of success of different attack actions.

\subsection{RT Server}
Table \ref{tab:RTSRV-PenTest} summarizes the penetration testing results for the attack tree in Figure \ref{fig:RTSRV-Attack-Tree}. Bypass mysql authentication attack failed due to enforced authentication configuration. Brute-force mysql login attack succeeded to obtain mysql weak root password "sql2563". With mysql root password, Linux password hash file was loaded successfully into mysql using LOAD\_FILE command and dumped for password cracking. Finally, Linux hashed password for root was recovered. The RT server is now under full control using root password and either SSH or remote desktop login. Attack injection is semi-automated using Metasploit modules. 

From penetration testing results, it can be safely assumed that mysql login authentication is enforced, especially for industrial plants, therefore $c_{12}=0$. In addition, we can assume that only one remote access capability is allowed, either remote desktop or SSH, for operational purposes. We arbitrarily choose SSH hence $c_{16}=0$. Therefore, the probability of RT Server compromise reduces to:
\begin{align}
P[\texttt{Srv-Comp}]= c_{13}c_{14}c_{15}c_{17}
\label{eq:RTSRV-ATTACK-PROB}
\end{align}
The probability of such an sql attack could be made zero by either closing the SSH port or enabling OS security on mysql database. Both options may not be possible for operational purposes. However, the probability could be made arbitrarily small by enforcing strong password policy for both mysql database and Linux server. It should be noted, however, that the enforcement of a very strong password policy may lead to difficult passwords to remember, hence written down passwords that may be leaked.
 
\begin{table}[tb!]
	\begin{center}
    \setlength{\tabcolsep}{10pt}
    \begin{tabular}{ |p{1.8cm}|p{1cm}|p{1cm}|p{1.75cm}| } 
				\hline
				\textbf{Vulnerability} & \textbf{Exploit Tool} & \textbf{Exploit Result} & \textbf{Notes} \\ 
				\hline
				Bypass mysql Authentication & Metasploit & Fail & mysql server configured to enforce authentication \\ \hline
				Brute-force mysql login & Metasploit & Success & Weak mysql root password used \\ \hline
				Linux password hash dump & Metasploit & Success & Apparmor disabled for mysql \\ \hline
				Crack Linux hashed passwords & Metasploit & Success & root password recovered \\ \hline
		\end{tabular}
	\end{center}
	\caption{Real Time Server Penetration Testing Results}
	\label{tab:RTSRV-PenTest}
\end{table}

\subsection{BPCS}
Table \ref{tab:BPCS-PenTest} summarizes the penetration testing results for the attack tree in Figure \ref{fig:BPCS-Attack-Tree}. Modbus STOP attack failed as it is not supported by the controller Modbus implementation. SYN flood attack did not also work because the BPCS controller runs RT Linux OS, which gives minimum guarantees on the control algorithm task periodicity. The Brute-force SSH login attack was successful in acquiring the root password, but unsuccessful in causing a DoS attack as the fail safe output is correctly configured on the controller. However, with controller root password, it is possible to overwrite the control program with malicious code and cause integrity attack. Modbus write attack was successful in writing the required value to the pre-determined register but the written value did not persist due to the periodic HMI-BPCS communication that continuously writes to the same register. Finally, random write to a contiguous address range was unsuccessful to cause a hazard without sufficient knowledge about the output-register mapping.

From the penetration testing results, we can make some simplifying assumptions to the attack tree probabilities. The probability of incorrect configuration for fail safe output could be set to 0, assuming the configuration is made once per controller software lifetime (we ignore here the possibility of a human error during controller reprogramming). Also, the probability that a random Modbus write would result in a hazard is very small given the large number of registers typically used in an industrial setup, so $a_2=0$. Finally, strong password policy enforcement decreases the probablity of a successful brute force SSH attack significantly, hence $c_4=0$. Accordingly, the probability of BPCS compromise reduces to:

\begin{align}
P[\texttt{BPCS-Comp}] = & c_8 (c_5 + c_6)
\label{eq:BPCS-ATTACK-Prob}
\end{align}

\begin{table*}[tb!]
	\begin{center}
    \begin{small}
    \setlength{\tabcolsep}{10pt}
    \begin{tabular}{ |p{2.5cm}|p{2cm}|p{1.5cm}|p{1.5cm}|p{6cm}| } 
				\hline
				\textbf{Vulnerability} & \textbf{Exploit Tool} & \textbf{Exploit Result} & \textbf{Hazard Caused} & \textbf{Notes} \\ 
				\hline
				Modbus STOP CPU Attack & Metasploit & Fail & No & Function not supported by BPCS Modbus implementation \\ \hline
				SYN Flood Attack to ports 22, 502 & Metasploit & Fail & No & RTOS does not degrade control algorithm performance with communication spikes \\ \hline
				Brute-force SSH login & Metasploit & Success & Y/N & user: admin, pass: "niroot" successfully deduced. DoS attack was unsuccessful due to fail safe setting. Integrity attack succeeded with proper software tools \\ \hline
				Modbus write to contiguous address range & Metasploit & Success & No & Random write to Modbus registers did not cause a process hazard \\ \hline
				Modbus write to a specific register & Metasploit & Success & No & HMI has periodic communication with BPCS that overwrites written data every 5 sec. \\ \hline
		\end{tabular}
		\end{small}
	\end{center}
	\caption{BPCS Penetration Testing Results}
	\label{tab:BPCS-PenTest}
\end{table*}

\subsection{HMI}
Table \ref{tab:HMI-PenTest} summarizes the penetration testing results for the attack tree in Figure \ref{fig:HMI-Attack-Tree}. RDP brute-force password attack was successful with user: operator and password: "reactorws". Such passwords are not uncommon in industrial plants as often times there is a repeated naming convention used, which is composed of plant unit name and WS for WorkStation. With remote access granted, access to HMI to switch reactor conroller to manual mode and close the coolant valve was possible. The Modbus attack was also successful in writing to Modbus registers at the HMI. However, the period of writing data has to be much shorter than both the Modbus poll delay between the HMI and BPCS and the GUI-Modbus registers writing loop frequency. This is mainly to avoid malicious data overwrite. This later Modbus attack assumes sufficient process knowledge and Modbus configuration data.

\begin{table*}[tb!]
	\begin{center}
		\begin{small}
			\setlength{\tabcolsep}{10pt}
			\begin{tabular}{ |p{2.5cm}|p{2cm}|p{1.5cm}|p{1.5cm}|p{6cm}| } 
				\hline
				\textbf{Vulnerability} & \textbf{Exploit Tool} & \textbf{Exploit Result} & \textbf{Hazard Caused} & \textbf{Notes} \\ 
				\hline
				Brute-force RDP login & Hydra & Success & Yes & user: admin, pass: "reactorws" successfully deduced. HMI access gained \\ \hline
				Modbus write to contiguous address range & Metasploit & Success & Yes & Process and Modbus configuration knowledge assumed \\ \hline
			\end{tabular}
		\end{small}
	\end{center}
	\caption{HMI Penetration Testing Results}
	\label{tab:HMI-PenTest}
\end{table*}

One finding from penetration testing that could simplify the probability of HMI-BPCS compromise in (\ref{eq:HMI-BPCS-Attack}) is that RDP lockout policy is an easy fix that could be configured in Windows 10 settings. Therefore, setting $c_{20}=0$ in (\ref{eq:HMI-BPCS-Attack}) we get:
\begin{align}
P[\texttt{HMI-BPCS}] = & c_5 c_7 + c_{18}c_{21}
\label{eq:HMI-BPCS-Attack-Prob}
\end{align}

\subsection{Overall Risk Assessment}
To assess the overall risk, a formula for risk scoring in terms of pre-defined risk metrics is required. This formula usually gives different weights for different risk metrics according to the organizational policy. The two key metrics in any risk assessment are the threat likelihood, $L$, and the cost of the consequences, $q$. In industry, and to simplify the analysis, the risk scoring function is usually categorical, using \emph{risk ranking} tables. The risk ranking table is on a matrix form where the rows represent finite likelihood categories and the columns represent finite consequence categories. The row-column intersection represent the risk rank, e.g., No, Low, Medium, and High risk. According to the risk rank, a Target Mitigated Event Likelihood (TMEL) is defined. If the event likelihood is > TMEL, then an additional protection layer is required to reduce the likelihood to the TMEL. More information on Layer Of Protection Analysis (LOPA) could be found in \cite{IEC61511}.

In this work, we adopt a continuous (rather than categorical) function that directly calculates the risk score. We define four risk metrics: Likelihood $L$, and three consequences; Safety loss $S$, Financial loss $F$, and Environmental loss $E$. We designate the target corporation risk score to be $r$. We define $q$ to be the normalized cost given by:
\begin{align}     \label{eq:Normalized Cost}
    & q = \alpha \left( \frac{S}{S_m} \right) + \beta \left( \frac{F}{F_m} \right) + \gamma \left( \frac{E}{E_m} \right) \\
    1 &= \alpha + \beta + \gamma \nonumber
\end{align}
$S_m,F_m,E_m$ are normalization factors representing the maximum value set for each relevant category, and $\alpha, \beta$ and $\gamma$ are weight factors defining the contribution of each metric to the overall risk score, and usually defined by the organization. The total cost $Q$ is a random variable defined on the sample space $\Omega = \{0,1,2,\ldots\}$ representing the number of hazardous events in a time interval. The best well-known risk score is given by the expected value of the total cost:
\begin{align}
    R = E[Q] = E[qN] = q E[N] = qL
\end{align}
where $L$ is the likelihood of the hazardous event in terms of the number of events per time interval. The target risk score $r$ is chosen according to the consequence. We assume the following linear relationship:
\begin{align}
    \log r = -\zeta q
    \label{eq:risk_score_q}
\end{align}
where $\zeta$ is a proportionality constant. The risk score is given by:
\begin{align}
    R = L \left[ \alpha \left( \frac{S}{S_m} \right) + \beta \left( \frac{F}{F_m} \right) + \gamma \left( \frac{E}{E_m} \right) \right]
\end{align}

The risk score has to satisfy $R \leq r$. In addition, since the likelihood is usually very small, we adopt the log function for the risk:
\begin{align}
    \log L &\leq \log r - \log \left[ \alpha \left( \frac{S}{S_m} \right) + \beta \left( \frac{F}{F_m} \right) + \gamma \left( \frac{E}{E_m} \right) \right] \nonumber \\
    &= -(\zeta q + \log q)
    \label{eq:Likelihood Constraint}
\end{align}
To calculate the likelihood $L$, we combine (\ref{eq:RTSRV-ATTACK-PROB}), (\ref{eq:BPCS-ATTACK-Prob}), and (\ref{eq:HMI-BPCS-Attack-Prob}) according to the attack tree in Figure \ref{fig:Reactor-Runaway-Attack-Tree-Abstract}. Ignoring higher order probability terms, the probability of reactor runaway is given by:
\begin{align}
P[\texttt{Runaway}] = c_{13}c_{14}c_{15}c_{17} \left[ c_8(c_5 + c_6) + c_5c_7 + c_8c_{21}\right] 
\label{eq:Prob-RUNAWAY}
\end{align}
where, without loss of generality, we assume a unity attack event per the chosen time interval. Assuming the probability that an attacker could compromise a corporate PC is given by $P_c$, the overall probability of reactor runaway is given by:
\begin{align}
L &= P_c P[\texttt{Runaway}] \nonumber \\
&= P_c c_{13}c_{14}c_{15}c_{17} \left[ c_8(c_5 + c_6) + c_5c_7 + c_8c_{21}\right] 
\label{eq:CPS-Likelihood}
\end{align}
Combining (\ref{eq:Likelihood Constraint}) and (\ref{eq:CPS-Likelihood}):
\begin{equation}
\begin{split}
    & P_c c_{13}c_{14}c_{15}c_{17} \left[ c_8(c_5 + c_6) + c_5c_7 + c_8c_{21}\right] \leq 10^{-(\zeta q + \log q)}
\end{split}
\label{eq:RunawayProbCondition}
\end{equation}
Equation (\ref{eq:RunawayProbCondition}) represents the design constraint for CPS security. Table \ref{tab:CPS-Design-Var} enumerates the design variables with description and possible countermeasures to reduce or eliminate the associated probabilities. It should be highlighted that some vulnerabilities could be entirely eliminated by taking the extreme policy of blocking the relevant services. However, the price paid is less flexibility in asset management and potentially higher cost of ownership.

\begin{table*}[tb!]
	\begin{center}
    \begin{small}
    \setlength{\tabcolsep}{10pt}
    \begin{tabular}{ |p{1.5cm}|p{5cm}|p{4cm}|p{3.5cm}| } 
				\hline
				\textbf{Design Variable} & \textbf{Description} & \textbf{Possible Countermeasures} & \textbf{Drawbacks} \\ 
				\hline
				$P_c$ & Probability of compromising a corporate PC & IT security policy  \\ \hline
				$c_5$ & Probability of leaking I/O and Modbus configuration & Non-technical - HR Policy \\ \hline
				$c_6$ & Probability of storing configuration documentation on the controller & Prevent local storage & Local storage is more convenient and guarantees no data loss \\ \hline
				$c_7$ & Probability of leaking process information & Non-technical - HR Policy  \\ \hline
				$c_8$ & Probability that HMI-BPCS comm. period is greater than reactor hazard time & Very low probability except for small reactors-Increase HMI-BPCS comm. frequency \\ \hline
				$c_{13}$ & Probability of successful brute-force mysql login attack & Enforce strong password policy & Forgetting passwords and writing them down \\ \hline
				$c_{14}$ & Probability of successful password recovery from a hash dump & Enforce strong password policy & Forgetting passwords and writing them down \\ \hline
				$c_{15}$ & Probability that security monitoring app is disabled for mysql & Increase security armoring for mysql application & Less flexibility in database management and interaction \\ \hline
				$c_{17}$ & Probability that SSH is enabled for RT Server & Block remote access & Operational inflexibility and higher operational cost \\ \hline
				$c_{18}$ & Probability of leaking operation passwords & Non-technical - HR Policy \\ \hline
				$c_{21}$ & Probability that RDP is enabled for HMI workstations & Disable RDP & Inflexible operational environment and higher HMI configuration cost \\ \hline
		\end{tabular}
		\end{small}
	\end{center}
	\caption{CPS Security Design Variables and Countermeasures}
	\label{tab:CPS-Design-Var}
\end{table*}

\subsection{Risk Assessment Results}
We apply the risk assessment methodology to the case study presented in the paper. It should be highlighted that the risk assessment methodology works by defining a target tolerable risk level, then working backwards to find an upper bound on the probability of cyber failures for CPS components. This is in comparison to the forward risk assessment process where the risk is assessed for a given system using its actual failure probability figures. The advantage of the backward approach is that it represents a proactive approach that is suitable for design time as well as runtime.

\begin{figure}[tb]
	\centering
	\includegraphics[scale = 0.15, trim = {4.2cm 0cm 0.6cm 0.6cm}, clip, angle=0]{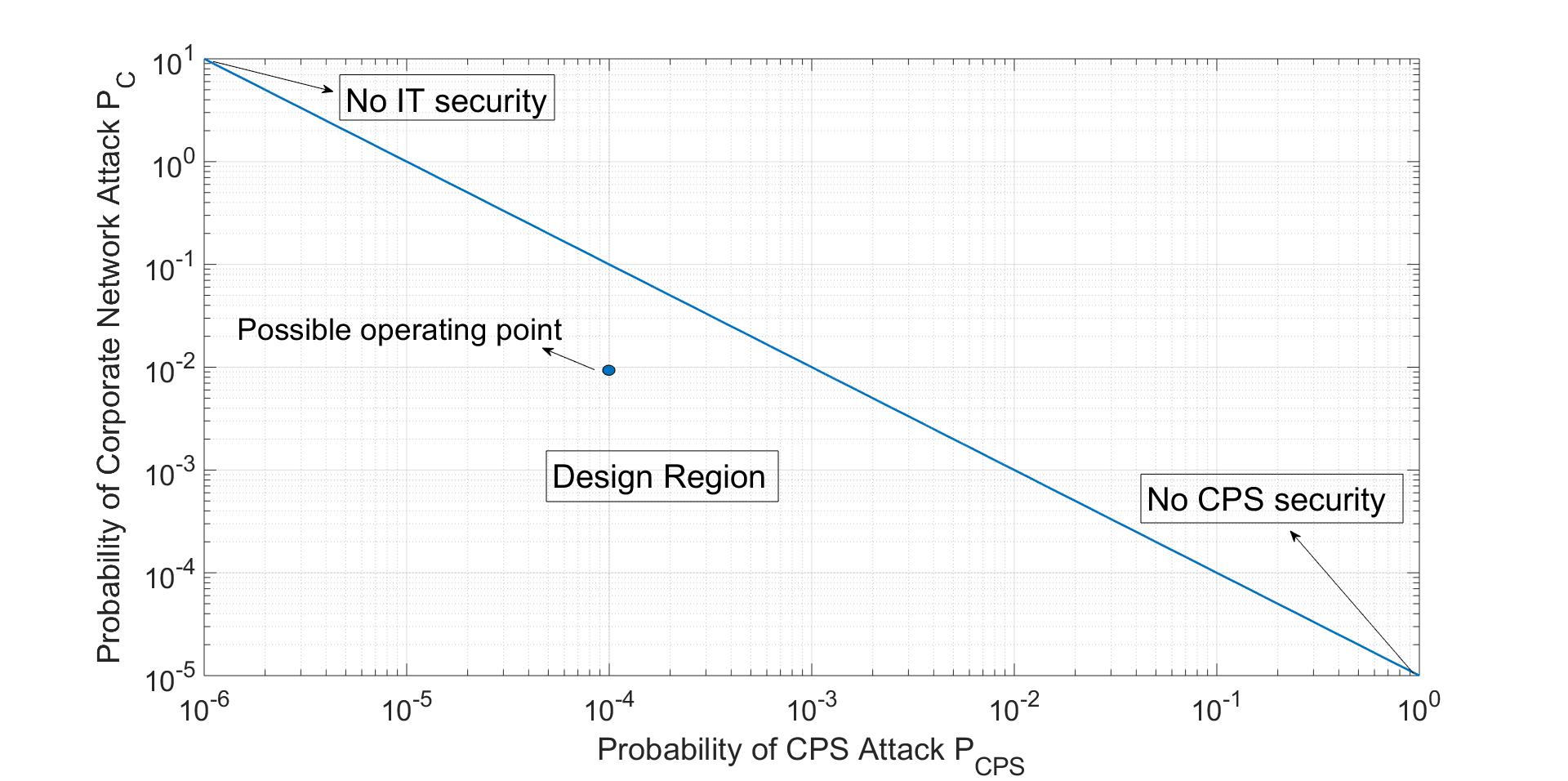}
	\caption{Cyber Security Design Curve. Any point in the area under the curve represents a valid design that satisfies the target mitigated risk level}
	\label{fig:Designcurve}
\end{figure}

Equation (\ref{eq:RunawayProbCondition}) represents the design equation for the system, where the exponent on the right hand side is calculated from (\ref{eq:Likelihood Constraint}) using the desired target risk level. All probabilities are considered design variables that need to be specified to achieve the target risk level. To illustrate the process, Table \ref{tab:Runaway-Design-Example} shows an instance of the design parameters and risk target for the reactor runaway risk scenario. The values in Table \ref{tab:Runaway-Design-Example} are used in (\ref{eq:Normalized Cost}) to yield a normalized cost $q=0.8005$. With $\zeta=6$, the target risk score is obtained from (\ref{eq:risk_score_q}) as $r \simeq 10^{-5}$. With these values, we can substitute in (\ref{eq:RunawayProbCondition}) to obtain the relationship between the design variables.
\begin{align}
P_c c_{13}c_{14}c_{15}c_{17} \left[ c_8(c_5 + c_6) + c_5c_7 + c_8c_{21}\right] \leq 10^{-5}
	\label{eq:designregion}
\end{align}
To get more insight into the design process, we make some assumptions regarding design variables. First, we assume that the probability of leaking information is the same whether it is software configuration or process documentation. Therefore, $c_5 = c_7$. Second, we assume the same password policy is enforced for different systems, including OS, database server, and remote desktop connection. Therefore, $c_{13}=c_{14}=c_{18}$. Third, the probability that HMI-BPCS communication period is greater than reactor hazard time is negligibly small for all practical purposes, so $c_8=0$. Fourth, since we chose SSH as the only remote configuration tool for the RT server and disabled remote desktop capability, we can set $c_{17}=1$. Finally, we assume remote desktop is disabled for all operator HMIs despite the operational inconvenience, hence $c_{21}=0$. This leads to the reduced design equation:
\begin{align}
\underbrace{(c^2_5c^2_{13}c_{15})}_{P_{\texttt{CPS}}} P_c \leq 10^{-5}
\label{eq:simplified-designregion}
\end{align}
where the first term $P_{\texttt{CPS}}$ represents the cyber attack failure probability due to CPS security weaknesses and $P_c$ represents the cyber attack failure probability due to IT security weaknesses. Equation (\ref{eq:simplified-designregion}) formalizes the interplay between IT security, as ending at the corporate network level, and CPS security, from the isolating firewall down to the plant floor. Figure \ref{fig:Designcurve} is a log-log plot for (\ref{eq:simplified-designregion}), where the design space is the area under the curve. Any point outside the design region results in a higher risk level than the tolerable value specified by the corporation. In addition, as one probability increases, the other probability has to decrease to compensate and achieve the target risk level. Examples are the extreme points (1,$10^{-5}$) and ($10^{-5}$,1). In practice, an operating point that represents a compromise between IT security and CPS security should be selected, e.g., ($P_{\text{CPS}},P_c$) = ($10^{-4},10^{-2}$) in Figure \ref{fig:Designcurve}. If the design constraint cannot be achieved, then a process modification, e.g., a safety instrumented function, needs to be added. This example shows the interplay between security and safety assessment, and that they cannot be carried out sequentially, but rather in an integrated way. This is one of the main insights gained from this work, which is further discussed in Section \ref{sec:discussion}.

\begin{table}[tb!]
	\begin{center}
    \begin{small}
    \setlength{\tabcolsep}{10pt}
    \begin{tabular}{ |p{1.5cm}|p{3cm}|p{1cm}|}
				\hline
				\textbf{Design Parameter} & \textbf{Description} & \textbf{Value} \\ 
				\hline
				$\alpha$ & Safety loss weight & 0.7 \\ \hline
				$\beta$ & Financial loss weight & 0.2 \\ \hline
				$\gamma$ & Environmental loss weight & 0.1 \\ \hline
				$S$ & Safety loss & 10 fatalities \\ \hline
				$S_m$ & Maximum safety loss & 10 fatalities \\ \hline
				$F$ & Financial loss & $\$5$ M \\ \hline
				$F_m$ & Maximum financial loss & $\$10$ M \\ \hline
				$E$ & Environmental loss & $\$10,000$ \\ \hline
				$E_m$ & Maximum environmental loss & $\$2$ M \\ \hline
				$ \zeta$ & Target risk prop. const & 6 \\ \hline
		\end{tabular}
		\end{small}
	\end{center}
	\caption{Reactor Runaway Risk Scenario - Design Example}
	\label{tab:Runaway-Design-Example}
\end{table}

\subsection{Hazard Mitigation}
The purpose of the risk assessment and countermeasure design is to minimize the probability of an attack that would cause a process hazard to the target mitigated likelihood. However, there is still a probability that an attack will be launched that cause a process hazard. In such case, a hazard mitigation strategy is mandatory.

One of the advantageous of the process hybrid automaton in Figure \ref{fig:Hybrid Automaton} is that it shows the best course of action when being in a hazardous state. The guard conditions on the outward transitions identify the process manipulation required to get out of the hazardous state. As an example, If the current process state is $S_4$, and a hazard is developed, then opening valve CV-2 will cause a transition to the safe state $S_1$. However, two key questions are in order: \begin{inparaenum}[(1)] \item how the current process state could be estimated?, and \item how can we take an action if the cyber system is compromised? \end{inparaenum} The first question concerns the problem of state estimation in hybrid systems which has been studied extensively in the literature \cite{Prakash2010}. To answer the second question, taking an action with a compromised cyber system has one of two ways; either having a backup image of the whole system that could be restored, or diagnosing the system online to isolate the attack and gaining control over the CPS. In both cases, the time taken to implement the mitigation action, $\tau_m$, should satisfy:

\begin{align}
	\tau_m < \tau_p +\tau_s - \tau_d
\end{align}
where $\tau_s$ is the time the system can operate outside its safe operating envelope with no damage and $\tau_d$ is the time taken to detect the hazard and $\tau_p$ is the process hazard development time, as reported in the system hybrid automaton. If this condition is not satisfied, e.g., covert attack that misleads the operator via malicious HMI data and resulting in high detection time $\tau_d$, then a process hazard will take place, and the mitigation action may or may not reduce the damage according to the process design. In such cases, a non-cyber (usually mechanical) mitigating solution has to be implemented in the process. For the reactor system, control valve CV-2 could have a manual override in the field to open or close it by a manual action. 

\section{Discussion} \label{sec:discussion}
\begin{figure}[tb]
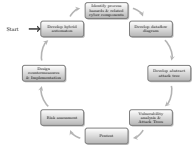

	\centering
	\resizebox{0.9\linewidth}{!}{%
	\smartdiagramset{uniform color list=gray!60 for all items, circular distance=4cm,text width=2.5cm,font=\scriptsize,
	additions={
   additional item offset=7.5mm,
   additional arrow color=gray!50!black,
 }
	}
	\smartdiagramadd[circular diagram:clockwise]{Develop hybrid automaton, Identify process hazards \& related cyber components, Develop dataflow diagram, Develop abstract attack tree, Vulnerability analysis \& Attack Trees, Pentest, Risk assessment, Design countermeasures \& Implementation}{left of module1/Start}
\smartdiagramconnect{-to}{additional-module1/module1}}
	\caption{Model-Based Design Cycle, as followed in the case study}
	\label{fig:Model-Based Design Cycle}
\end{figure}

The design and analysis process as applied to the case study in this paper could be generalized as in Figure \ref{fig:Model-Based Design Cycle}, where the physical process is included in the design cycle and hence subject to several iterations. A truly integrated approach to design both the safety and security systems should include both the physical and cyber systems at early stages of the design process.

The physical process modeling using hybrid automaton revealed several important insights. First, careful process modeling is crucial to identify the true hazards and associated cyber components to design a fit-for-purpose security system. Second, the time to develop a process hazard is an important parameter that should be taken into account during risk assessment and mitigation design. Along with the detection system, this may lead to more cost-effective cyber designs. Finally, as hazard generation is dependent on both the sequence of attack and time spent in each state, combining the attack traces from the hybrid automaton with the attacker probabilistic model results in more accurate risk assessment.

There is a tight coupling between safety and security, in terms of their impact on the physical system. Designing both systems independently may not lead to an optimal design. As the case study shows, only one physical component and two cyber nodes are the most critical components. All other components play a secondary role.

The model-based design approach followed in this work, where physical system modeling, data flow modeling, and attack trees are integrated provides one unified framework to design safe and secure CPS. The adoption of this integrated approach in industry is contingent on the development of software tools that automate most of the tasks that have been carried out manually in this work. This includes model development, hazard identification, data flow graph development, vulnerability scanning, attack tree development, penetration testing, and countermeasure design. Although several tools exist to automate individual tasks, an integrated tool that defines and implements the interface between different tasks is necessary.

The value of working backwards by identifying first the process hazards, though reducing significantly the number of attack scenarios, may not be readily obvious in centralized architectures. For example, the work presented shows that valve CV-2 is the sole critical component, and therefore, BPCS and HMI are the cyber components that should be protected. However, BPCS is the main process controller that implements all control algorithms, so the identification of CV-2 did not lead to effort or cost savings as the whole BPCS will be hardened anyway. For centralized systems like the CPS presented in this paper, this may lead to program segmentation and special protection for CV-2 related software modules. However, the significant value of this approach is revealed when considering distributed systems, where smart sensors and actuators are embedded devices that implement their own software and communicate over a common bus, without a need for a centralized controller.

\begin{figure}
    \centering
    \includegraphics[scale=0.55, trim = {0.5cm 0cm 0.6cm 1cm}, clip]{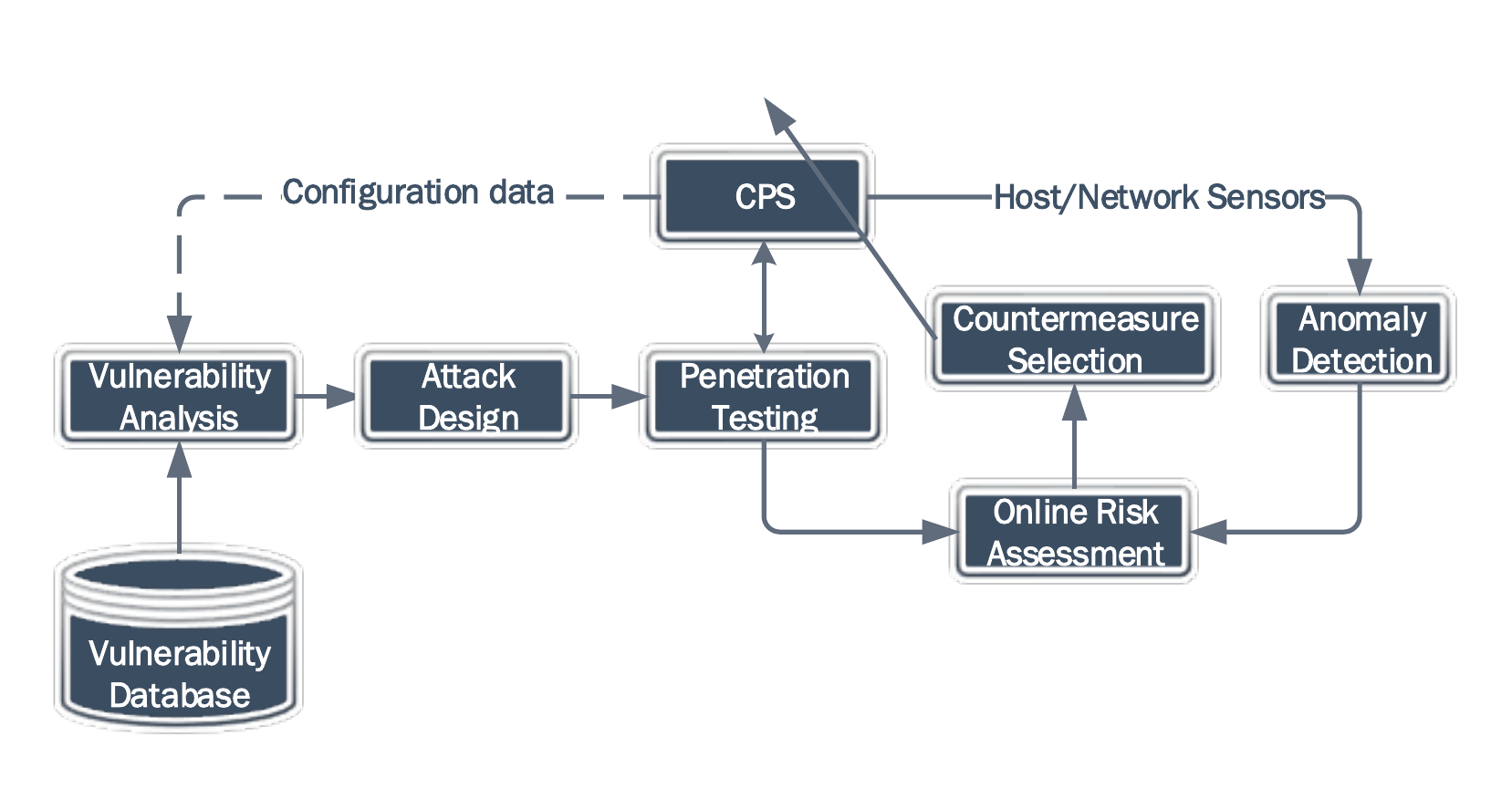}
    \caption{Model-Based risk assessment with day zero attacks}
    \label{fig:MBD-ML}
\end{figure}

Finally, it should be highlighted that the presented model-based design approach works for both offline and online risk assessment. However, for online risk assessment, it is also desirable to detect day zero attacks that are not available in vulnerability databases. In such case, the presented framework needs to be augmented with online anomaly detection to perform a complete risk assessment. The architecture of this solution is shown in Figure \ref{fig:MBD-ML}, including the possible countermeasure design. Figure \ref{fig:MBD-ML} represents a natural extension to the work presented in this paper.

\section{Conclusion and Future Work} \label{sec:Conclusion}
In this paper, we pursued an integrated approach to design the security system of a given CPS. The key finding is that by exploring the physical system behavior, it was observed that not all attacks could cause a system hazard, a hazard may take time to develop, and it may be possible to nullify the attack effect. Therefore, the paper highlights the need for an integrated approach to design the safety and security systems. Model-based design is a cornerstone for this integrated approach to be successful. For successful industrial adoption, a design automation tool chain that integrates physical and cyber domain modeling, as well as attack modeling and penetration testing, needs to be developed.

Several research directions could be identified from this work. First, the attacker profile was ignored in the risk assessment (all probabilities were assumed to be 1, certain events). This may lead to non-optimal system designs. Attackers posses different knowledge and skill set, and this needs to be captured in the risk assessment process. In addition, the probability that a process hazard may not be produced even though a process upset is caused due to random attacker actions was not considered. This factor may significantly decrease the overall likelihood of a process hazard post a cyber attack. Second, most successful attacks in the CPS domain are covert attacks that deceive the user via concurrent HMI manipulation. These attacks have complex structures and multiple objectives that need to be studied in more depth. Third, cyber attacks may lead to a system malfunction, and therefore should be included in the safety risk assessment as an initiating cause. This may lead to a reformulation of the current industrial practice for hazard identification and protection studies. Finally, when simulating the physical process for different cyber attack actions, it was assumed that the system disturbances are within operating limits. This assumption may be violated if the attack is organized against multiple system units simultaneously, resulting in the system being subject to concurrent disturbances and a cyber attack. The study of the composition of component risk assessment to yield an overall system risk assessment measure is an interesting research direction. Finally, the risk assessment process presented in this work is model-based, hence relying on known vulnerabilities and attack scenarios. However, day zero attacks represent a significant threat and challenge to CPS security. The integration of the model-based approach and machine learning approaches used for day zero attack detection in the context of online risk assessment is an important future research direction.

\nomenclature[A]{CPS}{Cyber-Physical System}
\nomenclature[A]{CSTR}{Continuous Stirred Tank Reactor}
\nomenclature[A]{SCADA}{Supervisory Control And Data Acquisition}
\nomenclature[A]{WSN}{Wireless Sensor Network}
\nomenclature[A]{BPCS}{Basic Process Control System}
\nomenclature[A]{SIS}{Safety Instrumented System}
\nomenclature[A]{HMI}{Human Machine Interface}
\nomenclature[A]{DMZ}{DeMilitarized Zone}
\nomenclature[A]{HAZOP}{HAZard and OPerability}
\nomenclature[A]{SIF}{Safety Instrumented Function}
\nomenclature[A]{IPL}{Independent Protection Layer}
\nomenclature[A]{GUI}{Graphical User Interface}
\nomenclature[A]{P\&ID}{Process \& Instrumentation Diagram}
\nomenclature[A]{MITM}{Man In The Middle}
\nomenclature[A]{SSH}{Secure Socket Shell}
\nomenclature[A]{DoS}{Denial of Service}
\nomenclature[A]{PID}{Proportional Integral Derivative}
\nomenclature[A]{LOPA}{Layer Of Protection Analysis}
\nomenclature[A]{TMEL}{Target Mitigated Event Likelihood}
\nomenclature[SB]{$P_c$}{Probability of gaining basic access to control LAN}
\nomenclature[SB]{$c_1$}{Probability No max number of passwords is configured}
\nomenclature[SB]{$c_2$}{Probability of leaking confidential company information}
\nomenclature[SB]{$c_3$}{Probability of guessing the correct password}
\nomenclature[SB]{$c_4$}{Probability that I/O configuration is stored on the controller}
\nomenclature[SB]{$c_5$}{Probability of incorrect fail safe output configuration}
\nomenclature[SB]{$c_7$}{Probability of cracking operator/engineer password for HMI access}
\nomenclature[SB]{$a_7$}{Probability that a random packet manipulation causes a hazard}
\nomenclature[SB]{$\alpha$}{Safety loss weight}
\nomenclature[SB]{$\beta$}{Financial loss weight}
\nomenclature[SB]{$\gamma$}{Environmental loss weight}
\nomenclature[SB]{$S$}{Safety loss}
\nomenclature[SB]{$S_m$}{Maximum safety loss}
\nomenclature[SB]{$F$}{Financial loss}
\nomenclature[SB]{$F_m$}{Maximum financial loss}
\nomenclature[SB]{$E$}{Environmental loss}
\nomenclature[SB]{$E_m$}{Maximum environmental loss}
\nomenclature[SB]{$\zeta$}{Target risk proportionality constant}
\printnomenclature

\begin{table*}[tb!]
	\begin{center}
	\begin{small}
    \setlength{\tabcolsep}{10pt}
    \begin{tabular}{ |p{1.5cm}|p{1.25cm}|p{4cm}|p{2cm}|p{5cm}|} 
			\hline
			\textbf{Node} & \textbf{Tool} & \textbf{Module} & \textbf{Purpose} & \textbf{Commands} \\ 
			\hline
			RT Server & nmap & - & Scan & \texttt{nmap -p- -sT -sV 192.170.1.2} \\ \hline
			RT Server & Metasploit & mysql\_authbypass\_hashdump & Bypass auth. & \texttt{set RHOSTS 192.170.1.2, set THREADS 50, exploit} \\ \hline
			RT Server & Metasploit & mysql\_login & Brute-force SQL login & \texttt{set RHOSTS 192.170.1.2, set USERNAME root, set PASS\_FILE ./passfile.lst, exploit} \\ \hline
			BPCS & nmap & - & Scan & \texttt{nmap -p- -sT -sV 192.168.1.10} \\ \hline
			BPCS & Metasploit & modicon\_command & Controller Remote Start/Stop & \texttt{set RHOSTS 192.168.1.10, set RPORT 502, set MODE STOP, exploit} \\ \hline
			BPCS & Metasploit & synflood & BPCS SYN Flood attack & \texttt{set RHOSTS 192.168.1.10, set RPORT 22, set RPORT 502, exploit} \\ \hline
			BPCS & Metasploit & ssh\_login & Brute-force SSH login & \texttt{set RHOSTS 192.168.1.10, set USERPASS\_FILE ./passfile.list, exploit} \\ \hline
			BPCS & Metasploit & modbusclient & Write to a single Modbus register & \texttt{set RHOSTS 192.168.1.10, set RPORT 502, set DATA\_ADDRESS 0, set DATA 0, exploit} \\ \hline
			BPCS & Metasploit & modbusclient & Write to multiple Modbus register & \texttt{set RHOSTS 192.168.1.10, set RPORT 502, set DATA\_ADDRESS 0, set DATA\_REGISTERS 0,100,100, exploit} \\ \hline
			HMI & nmap & - & Scan & \texttt{nmap -p- -sT -sV 192.168.1.20} \\ \hline
			HMI & Hydra & - & Brute force RDP & \texttt{Hydra –v –f -l operator -P ./passfile.list rdp://192.168.1.20} \\ \hline
			HMI & Metasploit & modbusclient & Write to multiple Modbus register & \texttt{set RHOSTS 192.168.1.20, set RPORT 502, set DATA\_ADDRESS 0, set DATA\_REGISTERS 0,100,100, exploit} \\ \hline
		\end{tabular}
		\end{small}
	\end{center}
	\caption{Vulnerability Scanning and Penetration Testing Commands}
	\label{tab:VULN-PENTEST-CMD}
\end{table*}

\section*{Acknowledgment}
This research was made possible by NPRP 9-005-1-002 grant from the Qatar National Research Fund (a member of The Qatar Foundation). The statements made herein are solely the responsibility of the authors.

\bibliographystyle{ieeetr}

\balance




\end{document}